\documentclass[aps,amsmath]{revtex4}
\usepackage{graphicx}% Include figure files

        \newcommand{\be}{\begin{equation}}
        \newcommand{\ee}{\end{equation}}
        \newcommand{\bea}{\begin{eqnarray}}
        \newcommand{\eea}{\end{eqnarray}}
        \newcommand{\ban}{\begin{eqnarray*}}
        \newcommand{\ean}{\end{eqnarray*}}
        \newcommand{\half}{\frac{1}{2}}

\begin{document}

\preprint{}

\title{Chiral symmetry restoration in linear sigma models with
different numbers of quark flavors}

\author{Dirk R\"oder}
\email{roeder@th.physik.uni-frankfurt.de}
\author{J\"org Ruppert}
\email{ruppert@th.physik.uni-frankfurt.de}
\author{Dirk H. Rischke}
\email{drischke@th.physik.uni-frankfurt.de}
\affiliation{Institut f\"ur Theoretische Physik,
Johann Wolfgang Goethe-Universit\"at \\
Robert-Mayer-Str.\ 8--10, D-60054 Frankfurt/Main, Germany \\
}

\begin{abstract}
Chiral symmetry restoration at nonzero temperature is studied
in the framework of the $O(4)$ linear sigma model and the $U(N_f)_{r} \times
U(N_f)_{\ell}$ linear sigma model with $N_f=2,\, 3$ and 4 quark flavors.
We investigate the temperature dependence of the masses
of the scalar and pseudoscalar mesons, and the non-strange, strange, 
and charm condensates within the Hartree
approximation as derived from the Cornwall-Jackiw-Tomboulis formalism.
We find that the masses of the non-strange and strange mesons at nonzero 
temperature depend sensitively on the particular 
symmetry of the model and the number of
light quark flavors $N_f$. On the other hand, due to the large
charm quark mass, neither do charmed mesons significantly affect the
properties of the other mesons, nor do their masses change appreciably
in the temperature range around the chiral symmetry restoration temperature.
In the chiral limit, the transition temperatures for chiral symmetry
restoration are surprisingly close to those found in lattice QCD.
\end{abstract}

\date{\today}% It is always \today, today,
             %  but any date may be explicitly specified
\pacs{11.10.Wx}% PACS, the Physics and Astronomy
                             % Classification Scheme.
%\keywords{Suggested keywords}%Use showkeys class option if keyword
                              %display desired
\maketitle

\section{Introduction} \label{I}

For $N_f$ massless quark flavors, the QCD Lagrangian has a chiral
$U(N_f)_r \times U(N_f)_{\ell} = SU(N_f)_r \times SU(N_f)_\ell
\times U(1)_V \times U(1)_A$ symmetry. 
Here, $V=r+\ell$, while $A= r - \ell$. 
The $U(1)_V$ symmetry corresponds to baryon number conservation.
It is always respected and thus plays no role in the symmetry breaking
patterns considered in the following. In the vacuum, 
a non-vanishing expectation value of the quark condensate,
$\langle \bar{q}_{\ell} \, q_{r} \rangle \neq 0$, spontaneously
breaks the above symmetry to $SU(N_f)_{V}$.
This gives rise to $N_f^2$ Goldstone bosons which dominate
the low-energy dynamics of the theory.
As shown by 't Hooft \cite{'tHooft:1976fv,'tHooft:1976up},
instantons break the $U(1)_{A}$ symmetry explicitly to
$Z(N_{f})_{A}$ \cite{Pisarski:1984ms}. (For the low-energy dynamics
of QCD, however, this discrete symmetry is irrelevant.)
Consequently, one of the $N_f^2$ Goldstone bosons becomes massive, leaving
$N_f^2-1$ Goldstone bosons.
The $SU(N_f)_r \times SU(N_f)_{\ell} \times U(1)_A$ symmetry of the 
QCD Lagrangian is also explicitly broken
by nonzero quark masses. The $N_f^2-1$ low-energy degrees of freedom 
then become pseudo-Goldstone bosons. For $M \leq N_{f}$ degenerate quark
flavors, an $SU(M)_{V}$ symmetry is preserved.

As indicated by lattice QCD calculations \cite{Karsch:2001cy}, 
chiral symmetry is restored at temperatures around $\sim 150$ MeV 
(at zero net-baryon number density). 
It is difficult to determine the order of the chiral
phase transition on the lattice. At this time, lattice calculations have not
unambiguously answered this question. For physical values of the 
quark masses, calculations with staggered fermions \cite{Brown:1990ev}
favor a smooth crossover transition, while calculations with Wilson fermions
\cite{Iwasaki:1996yj} predict the transition to be of first order.

For vanishing quark masses, i.e., in the chiral limit, however, one
can use universality arguments to determine the order of the
phase transition. According to universality,
the order of the chiral transition in QCD is identical to that
in a theory with the same chiral symmetries as QCD, for instance, 
the $U(N_f)_r \times U(N_f)_\ell$ linear sigma model.
This argument was employed by
Pisarski and Wilczek \cite{Pisarski:1984ms} who found that
for $N_f=2$ flavors of massless quarks, the transition
can be of second order, if the $U(1)_A$ symmetry is explicitly
broken by instantons. It is driven first order by fluctuations, if
the $U(1)_A$ symmetry is restored at $T_c$. For $N_f=3$ massless
flavors, the transition is always first order. In this case, the
term which breaks the $U(1)_A$ symmetry explicitly is a cubic
invariant, and consequently drives the transition first order. In
the absence of explicit $U(1)_A$ symmetry breaking, the transition
is fluctuation-induced of first order. For $N_f=4$, the same
argument leads to a first order chiral transition in the 
absence of the $U(1)_A$ anomaly. The term which breaks
the $U(1)_A$ symmetry is no longer a 
cubic invariant, but of quartic order in the fields.
Since this term does not generate an infrared-stable fix point, 
the transition remains of first order.

For nonzero quark masses, the chiral symmetry of QCD is explicitly 
broken. Nonzero quark masses act like a magnetic field in spin
systems, such that a second order phase transition becomes a
crossover transition. When the quark masses increase,
a first order phase transition may for a while remain of first order, but it
will ultimately become a crossover transition, too. 
In order to decide whether this happens for a particular choice
of quark masses, universality arguments cannot be applied and
one has to resort to numerical calculations.
As an alternative to lattice QCD calculations, one can also
use linear sigma models to investigate this question
\cite{Lenaghan:2000kr}. 

Studying the linear sigma model at nonzero temperature, however,
requires many-body resummation schemes, because infrared
divergences cause naive perturbation theory to break down
\cite{Dolan:1974qd}. In these resummation schemes, one necessarily
has to make approximations by selecting certain subsets of diagrams.
The most commonly employed scheme is the Hartree approximation.
This approximation fails to reproduce the
correct order of the chiral transition for the $O(4)$ linear sigma model:
the transition is of second order, while the Hartree approximation 
yields a first order transition.
For $U(N_f)_r \times U(N_f)_\ell$ linear sigma models with $N_f = 3$, 
however, the Hartree approximation correctly produces
a first order chiral transition \cite{Lenaghan:2000ey}. For $N_f=2$,
in the absence of the $U(1)_A$ anomaly the transition is of first
order and, as will be shown in this paper, the Hartree approximation 
agrees with this result. With the $U(1)_A$ anomaly, the transition is
of second order. We shall find that the Hartree approximation (slightly)
violates this prediction by producing a (weak) first order transition.

The utility of linear sigma models, however, transcends further
than just predicting the order of the chiral phase transition.
The degrees of freedom of the $U(N_f)_r \times U(N_f)_\ell$
linear sigma model are the scalar and pseudoscalar mesons.
In the vacuum, the latter are the (pseudo-) Goldstone bosons of 
chiral symmetry breaking. Thus, the linear sigma model can also be viewed
as an effective low-energy theory for QCD. 
At high temperatures, chiral symmetry is restored. Consequently, 
chiral partners among the scalar and pseudoscalar mesons must 
become degenerate in mass.
Since the $U(N_f)_r \times U(N_f)_\ell$ linear sigma model treats
both scalar and pseudoscalar mesons on the same footing, it is
particularly suited to describe the change of meson properties
across the chiral transition.
Note that, in the case of a first order chiral transition which coincides
with the deconfinement transition, the correct degrees of freedom
in the high-temperature phase are quark and gluons instead of mesons. 
However, in the case of a crossover transition mesonic degrees of
freedom are well-defined even above the chiral transition temperature.

\begin{table}
\begin{center}
\begin{tabular}{|l|c|c|c|c|}\hline
 &$O(4)$&$U(2)_{r}\times U(2)_{\ell}$ & $U(3)_{r}\times U(3)_{\ell}$ &
$U(4)_{r}\times U(4)_{\ell}$ \\ \hline
Explicit chiral        &&&&\\
symmetry breaking   &$\surd$&$\surd$&$\surd$&$\surd$\\
with $U(1)_A$        &&&&\\
anomaly           &&&&\\ \hline
Explicit chiral        &&&&\\
symmetry breaking   &$\times$&$\surd$&$\surd$& $ -$ \\
without $U(1)_A$           &&&&\\
anomaly            &&&&\\\hline
Chiral limit        &&&&\\
with $U(1)_A$   &$\surd$&$\surd$&$\surd$& $-$\\
anomaly  &&&&\\\hline
Chiral limit       &&&&\\
without $U(1)_A$ &$\times$&$\surd$&$\surd$& $-$\\
anomaly &&&&\\\hline
\end{tabular}
\vspace{3mm} 
\caption{The table shows the symmetry
breaking patterns studied in this paper. 
The $\surd$ symbol indicates that the specific
symmetry breaking pattern is physically interesting and investigated
here. The $\times$ symbol indicates that this symmetry breaking
pattern does not exist at all and the $-$ symbolizes that the
respective symmetry breaking pattern is not studied. }
\label{table1}
\end{center}
\end{table}

In the present work, we follow this line of arguments and
investigate the change of meson masses and
quark condensates with temperature 
in the framework of $U(N_f)_r \times U(N_f)_\ell$ linear sigma models.
We focus in particular on the question how the number of quark
flavors $N_f$ changes the temperature dependence of the meson masses
and the quark condensates.
For a given $N_f$, we furthermore investigate the 
different patterns of symmetry breaking
arising from the presence or absence of the $U(1)_A$ anomaly, and 
from taking zero or nonzero values for the quark masses.
In this sense, the present study is an extension of previous
work \cite{Lenaghan:1999si,Lenaghan:2000ey}.
Table \ref{table1} presents an overview of the different models
and patterns of symmetry breaking studied in this paper.

The $U(2)_r \times U(2)_\ell$ linear sigma model
has eight degrees of freedom: the scalar fields are the $\sigma$ meson 
and the three $a_0$ mesons, the pseudoscalar fields are the $\eta$ meson
and the three pions. With spontaneous breaking of the $U(2)_A$ chiral
symmetry, but without the $U(1)_A$ anomaly, 
the pions and the $\eta$ meson are (pseudo-) Goldstone
bosons, while the $\sigma$ and $a_0$ meson are heavy states.
With the $U(1)_A$ anomaly, the $\eta$ meson also becomes heavy. 

With explicit breaking of the $U(1)_A$ symmetry due to the anomaly
(and neglecting the $U(1)_V$ symmetry of baryon number conservation),
the remaining symmetry of the $U(2)_r \times U(2)_\ell$ linear sigma model is 
$SU(2)_r \times SU(2)_\ell$. This group is isomorphic to $O(4)$.
The appropriate effective theory incorporating this
symmetry is the $O(4)$ linear sigma model
\cite{Gell-Mann:1960np}. This model has only four degrees of freedom,
the $\sigma$ meson and the three pions. In this sense, it represents the
limit of the $U(2)_r \times U(2)_\ell$ model for
maximum $U(1)_A$ symmetry breaking. 
The $U(2)_r \times U(2)_\ell$ model is more general, since
it allows to consider the properties of mesons also 
without the $U(1)_A$ anomaly.

In the limit of maximum $U(1)_A$ symmetry breaking, 
the $\eta$ and $a_0$ mesons become infinitely heavy and
are thus removed from the spectrum of physical excitations.
This approximation is justified at small temperatures, 
where the dynamics is determined by the
lightest hadronic degrees of freedom, i.e., the pions and, to some extent, the
$\sigma$ meson. At higher temperatures and, in particular, around the
chiral transition, however, heavier mesons become more and more
important. It is therefore interesting to compare the results of the 
$O(4)$ model with those of the $U(2)_r \times U(2)_\ell$ model
which incorporates these heavier degrees of freedom. 
Note that the $O(4)$ linear sigma model at nonzero temperature has been
discussed in Refs.\ \cite{Petropoulos:1998gt,Lenaghan:1999si}.

In the physical hadron spectrum, the kaons are lighter than the
$a_0$ and $\eta$ mesons and thus are more copiously produced at
nonzero temperature. Consequently, the kaons are 
expected to influence the dynamics of the system to a larger extent
than the $a_0$ and $\eta$ mesons.
Therefore, the $U(2)_r \times U(2)_\ell$ model
is unrealistic in the sense that it neglects these strange meson
degrees of freedom. From a physical point of view, it is thus
necessary to enlarge the symmetry group to $U(3)_r \times U(3)_\ell$ and
study the corresponding linear sigma model.
Nevertheless, it is still interesting to compare 
the $U(2)_r \times U(2)_\ell$ model with 
the $U(3)_r \times U(3)_\ell$ model in order to see how 
the strange degrees of freedom affect the results.
The $U(3)_r \times U(3)_\ell$ model was previously studied at nonzero
temperature in Ref.\ \cite{Lenaghan:2000ey}.

Finally, we extend our investigations by including the charm degree
of freedom. In principle, the corresponding linear sigma model has an
$SU(4)_r \times SU(4)_\ell \times U(1)_A$ symmetry, 
but in nature this symmetry is strongly
explicitly broken by the large charm quark mass. Therefore, we only
consider the physically relevant
case of explicit chiral symmetry breaking with $U(1)_A$ anomaly.

For all cases considered (see Table \ref{table1}) we
study the meson masses and the condensates as functions of temperature. 
In the chiral limit, we determine the phase transition temperatures
by computing the effective potentials. In this aspect, we extend
the previous study of Ref.\ \cite{Lenaghan:2000ey}. For
the $O(4)$ model, this has already been done
in Ref.\ \cite{Petropoulos:1998gt}.

Our calculations are done in the Hartree approximation which we
derive within the Cornwall-Jackiw-Tomboulis (CJT) formalism 
\cite{Cornwall:1974vz}. The Hartree approximation 
has the advantage that the meson
self-energies become independent of momentum and energy and one
only has to solve gap equations for the meson masses as a function
of temperature. To go beyond the Hartree approximation, for instance by
including energy-momentum dependent contributions to the
self-energies, is considerably more difficult 
\cite{vanHees:2000bp,vanHees:2001ik,vanHees:2001pf,vanHees:2002bv},
and will be deferred to a future publication.

The remainder of this paper is organized as follows.  
In Sec.\ \ref{II}, we review the formulas of the
CJT formalism, which are relevant for the present study, and apply
them to the $O(4)$ model and the $U(N_f)_r \times U(N_f)_\ell$ linear
sigma models for $N_f=2,3,$ and 4.
In Sec.\ \ref{III}, it is shown how to determine the coupling 
constants of the
different models from the vacuum properties of the mesons and
condensates. In Sec.\ \ref{IV}, we discuss
the temperature dependence of the masses and the condensates.
We conclude this work in Sec.\ \ref{V} with a summary of our results.

We use the imaginary-time formalism to compute quantities at
nonzero temperature. Our notation is 
\be 
\int_k \, f(k) \equiv T \sum_{n=-\infty}^{\infty}
                        \int \frac{d^{3}{\bf k}}{(2\pi)^3} \,
         f(2 \pi i n T,{\bf k}) \,\,\,\, , \,\,\,\,
\int_x \, f(x) \equiv \int^{1/T}_{0} d \tau \int d^{3}{\bf x} \,
f(\tau,{\bf x}) \,\, . 
\ee 
Our units are $\hbar=c=k_{B}=1$. The metric tensor is 
$g^{\mu \nu} = {\rm diag}(+,-,-,-)$.  Throughout
this work, all latin subscripts are adjoint $U(N_f)$ indices,
$a=0,\ldots,N_f^2-1$, and a summation over repeated indices is
understood.

\section{Linear Sigma Models in the Hartree approximation} \label{II}

\subsection{The CJT formalism} \label{IIa}

The CJT formalism \cite{Cornwall:1974vz}
generalizes the concept of the effective action 
for one-point functions to that of an effective action
for one- and two-point functions. It is particularly useful
for theories with spontaneously broken symmetry. In this case,
truncating the standard loop expansion for the effective action 
\cite{Jackiw:1974cv} at some given order, at nonzero temperature
one may obtain unphysical, tachyonic propagation of quasiparticles 
with small momenta. 
The reason for this failure of the standard loop expansion
is that only the expectation value of the one-point function 
is self-consistently determined in this approach. The CJT formalism
goes beyond the standard loop expansion by self-consistently
determining the expectation value for the two-point function in
addition to that of the one-point function. Effectively, this amounts
to solving a Dyson-Schwinger equation for the two-point functions
and yields a self-consistent computation of the quasiparticle self-energy.
For translationally invariant systems, the effective action 
becomes the effective potential. To give an example,
consider the general Lagrangian
\be \label{L} 
{\cal L}(\varphi) = \frac{1}{2} \, \partial_\mu \varphi \,
\partial^\mu \varphi - U(\varphi) 
\ee
for a scalar quantum field $\varphi$.
The effective potential in the CJT formalism reads
\be \label{14}
V[\bar{\phi},\bar{G}] = U(\bar{\phi}) +\half \int_k \, \ln \bar{G}^{-1}(k) + 
\half \, \int_k\, \left[G^{-1}(k;\bar{\phi}) \,
\bar{G}(k) - 1 \right] + V_2[\bar{\phi},\bar{G}] \,\, , 
\ee
where $\bar{\phi}$ is a $c$-number field (it is the expectation
value of the quantum field $\varphi$ in the presence of an external source),
$U(\bar{\phi})$ is the classical potential energy density (the tree-level
potential) in the Lagrangian  (\ref{L}),
and $G^{-1}$ is the inverse of the tree-level propagator,
\be 
G^{-1} (k;\bar{\phi}) \equiv  -k^2 + U''(\bar{\phi})\,\,.
\ee
Here, $U''(\bar{\phi})$ is the second derivative of $U(\bar{\phi})$ 
with respect to $\bar{\phi}$. The last term in Eq.\ (\ref{14}), 
$V_2[\bar{\phi},\bar{G}]$, 
is the sum of all two-particle irreducible vacuum diagrams where all lines
represent full propagators $\bar{G}$.
The expectation values of the one-point function, $\phi$,
and of the two-point function, ${\cal G}(k)$ are determined from
the stationarity conditions
\begin{subequations}
\bea \label{stationphi} 
\left. \frac{\delta V[\bar{\phi},\bar{G}]}{\delta \bar{\phi}} \,
\right|_{\bar{\phi}=\phi,\, \bar{G}= {\cal G}}= 0 \,\, ,\\
\left. \frac{\delta V[\bar{\phi},\bar{G}]}{\delta \bar{G}(k)} \,
\right|_{\bar{\phi}=\phi,\, \bar{G}= {\cal G}}= 0 \,\, .
\label{stationG} 
\eea 
\end{subequations}
With Eq.\ (\ref{14}), the latter can be
written in the form 
\begin{subequations}
\be \label{111} 
{\cal G}^{-1}(k) = G^{-1}(k;\phi) + \Pi(k)\,\, , 
\ee
where 
\be \label{selfenergy} 
\Pi(k) \equiv 2 \left. \frac{\delta
V_2 [\bar{\phi},\bar{G}]}{\delta \bar{G}(k)} 
\right|_{\bar{\phi}=\phi,\, \bar{G}= {\cal G}} 
\ee 
\end{subequations}
is the self-energy. Since $\Pi(k)$ is in general a functional of
${\cal G}$, Eq.\ (\ref{111}) represents a Schwinger-Dyson equation for
the full (dressed) propagator.

The standard effective potential for the expectation value of the
one-point function, $V(\bar{\phi})$, 
is obtained from the effective potential (\ref{14})
by taking the full propagator $\bar{G}$ to be a function of $\bar{\phi}$, 
instead of an independent variable,
\be
V(\bar{\phi}) = V [\bar{\phi},\hat{G}(\bar{\phi})] \,\,,
\ee
where $\hat{G}(k;\bar{\phi})$ is determined from
\be 
\left. \frac{\delta V[\bar{\phi},\bar{G}]}{\delta \bar{G}(k)} \, 
\right|_{\bar{G} = \hat{G}} = 0 \,\, ,
\label{stationVeff}
\ee
which is equivalent to
\begin{subequations}
\be  
\hat{G}^{-1}(k;\bar{\phi}) = G^{-1}(k;\bar{\phi}) + 
\hat{\Pi}(k;\bar{\phi})\,\, ,  \label{hatG}
\ee
where 
\be \label{selfenergyeff} 
\hat{\Pi}(k;\bar{\phi}) \equiv 2 \left.
\frac{\delta V_2 [\bar{\phi},\bar{G}]}{\delta \bar{G}(k)} 
\right|_{ \bar{G}= \hat{G}}. 
\ee 
\end{subequations}
This expression and Eq.\ (\ref{14}) can be used
to obtain a compact form of the standard effective potential 
\be
V(\bar{\phi}) = 
U(\bar{\phi}) + \half \int_k \, \ln
\hat{G}^{-1}(k;\bar{\phi}) -  \, \frac{1}{2}\int_k \, 
\hat{\Pi}(k;\bar{\phi})\, 
\hat{G}(k;\bar{\phi}) +  V_2[\bar{\phi},\hat{G}] \,\,. \label{Veff} 
\ee
Since $V_2$ contains infinitely many diagrams, an exact calculation
is impossible. In practice, one has to restrict the
computation of $V_2$ to a finite number of diagrams. The selected
set of diagrams defines a particular many-body approximation. 
Cutting internal lines in these diagrams according 
to Eq.\ (\ref{selfenergy}), one obtains the diagrams contributing to
the self-energy of the quasiparticles in this approximation scheme. 
Solving the Dyson-Schwinger equation (\ref{111}) provides a self-consistent 
calculation of this self-energy. In general,
the Dyson-Schwinger equation is an integral equation for the
self-energy as a function of energy and momentum.

\begin{figure}
\includegraphics[height=8cm]{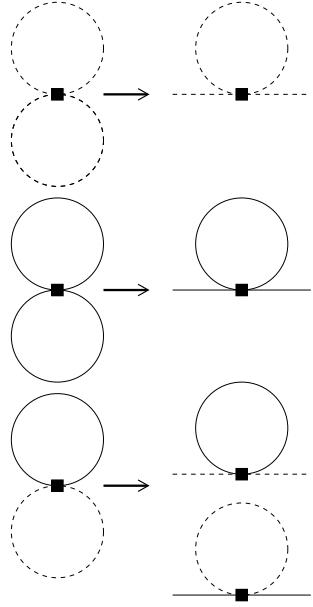}
\caption{Left-hand side: the double-bubble diagrams. Full lines are scalar
particles, dashed lines are pseudoscalar particles. 
Right-hand side: the tadpole contributions to the self-energies
obtained by cutting a line in the double-bubble diagrams on the 
left-hand side.}
\label{dbubble}
\end{figure}

If one only takes the double-bubble diagrams 
on the left-hand side of Fig.\ \ref{dbubble} into account in the 
calculation of $V_2$, one obtains 
the Hartree approximation. Cutting these diagrams
yields the well-known tadpole diagrams for the self-energy,
cf.\ the right-hand side of Fig.\ \ref{dbubble}.
The Dyson-Schwinger equation (\ref{111}) is a self-consistency 
equation for this self-energy due to the fact that the internal lines in the
tadpole diagrams represent full propagators.
The Hartree approximation is a particularly simple many-body approximation
scheme, because the tadpole diagrams are independent of energy and momentum,
and thus the Dyson-Schwinger equations are no longer integral equations,
but become fix-point equations for the quasiparticle masses.

\subsection{The $O(4)$ model}\label{IIb}

In this subsection, we apply the CJT formalism to the
$O(N)$ model. Our numerical results presented in 
Sec.\ \ref{IV} are exclusively for the case $N=4$.
The CJT effective potential for the $O(N)$ model is 
\cite{Lenaghan:1999si}
\bea
V[\bar{\sigma},\bar{S},\bar{P}]
 & = & U(\bar{\sigma})  +  \half \int_k \left[\, \ln \bar{S}^{-1}(k) +
        S^{-1}(k;\bar{\sigma})\, \bar{S}(k)-1 \, \right] \nonumber \\
 & + &  \frac{N-1}{2} \int_k \, \left[\,
        \ln \bar{P}^{-1}(k) + P^{-1}(k;\bar{\sigma})\, \bar{P}(k)-
        1 \, \right]  +  V_{2}[\bar{\sigma},\bar{S},\bar{P}]\,\, ,
\eea 
where $\bar{\sigma}$ is the expectation value of the scalar field
(in the presence of external sources).
Because the vacuum of QCD has even parity, the expectation values
of the pseudoscalar fields can be set to zero.
The quantities $\bar{S}$ and $\bar{P}$ are the full propagators for
scalar and pseudoscalar particles, while $S^{-1}$ and $P^{-1}$ are the
corresponding inverse tree-level propagators,
\begin{subequations}
\bea  \label{Dsigma} 
S^{-1}(k;\bar{\sigma}) &=& -k^{2} + m_\sigma^2(\bar{\sigma})
\,\, , \\
P^{-1}(k;\bar{\sigma}) &=& -k^{2} + m_\pi^2 (\bar{\sigma})\,\, , 
\label{Dpi}
\eea
\end{subequations}
where the tree-level masses are
\begin{subequations}
\bea
m_\sigma^2(\bar{\sigma}) & = & \mu^{2} + 
        \frac{12\, \lambda}{N}\, \bar{\sigma}^{2} \,\, , \\
m_\pi^2(\bar{\sigma}) & = &    \mu^2 +
        \frac{4\, \lambda}{N}\, \bar{\sigma}^{2} \,\, .
\eea
\end{subequations}
The constant $\mu^2$ is the bare mass term in the Lagrangian
of the $O(N)$ model, while $\lambda$ is the four-point coupling constant.
For $\mu^2 <0$, the $O(N)$ symmetry is spontaneously broken to $O(N-1)$,
leading to $N-1$ Goldstone bosons. The tree-level potential is
\be \label{UtreeON}
U(\bar{\sigma}) =  \half \, \mu^{2}\, \bar{\sigma}^{2} +
        \frac{ \lambda}{N}\, \bar{\sigma}^{4} - H \,\bar{\sigma}\,\, ,
\ee
where $H$ is a term which breaks the $O(N)$ symmetry explicitly
to $O(N-1)$.
$V_2[\bar{\sigma},\bar{S},\bar{P}]$ is the sum of all two-particle
irreducible diagrams. In the following, we restrict ourselves
to the Hartree approximation, i.e., we take into account only the
double-bubble diagrams shown on the left-hand side of Fig.\ \ref{dbubble}. 
These diagrams have no explicit $\bar{\sigma}$ dependence. 
Then, only tadpole diagrams (with resummed propagators) contribute to the 
self-energies, cf.\ the right-hand side of Fig.\ \ref{dbubble}. 
In the Hartree approximation, 
\bea
V_{2}[\bar{S},\bar{P}] &=& 3\, \frac{\lambda}{N}
        \left[ \int_k \, \bar{S}(k) \right]^{2} +
        (N+1)(N-1)\, \frac{\lambda}{N} \left[
        \int_k \, \bar{P}(k)\right]^{2}
\nonumber \\
   & + & 2\,(N-1) \frac{\lambda}{N}
        \left[\int_k\, \bar{S}(k) \right] \left[\int_p \,
        \bar{P}(p) \right] \,\, \label{V2ON}.
\eea 
The stationarity conditions (\ref{stationphi}) and (\ref{stationG}) read
\begin{subequations}
\bea  \label{stationONH} 
H & = &   \mu^2 \, \sigma + \frac{4 \lambda}{N}\, \sigma^3
 +  \frac{4\lambda}{N}\, \sigma \int_q\, 
\left[ 3 \,{\cal S}(q) + (N-1) \, {\cal P}(q)\right] \,\, ,  \\ 
\label{stationONS} 
{\cal S}^{-1}(k) & = & - k^2 + M_\sigma^2 \,\, , \\
\label{stationONP} 
{\cal P}^{-1}(k) & = & - k^2 + M_\pi^2 \,\, , 
\eea
\end{subequations}
where ${M}_{\sigma}$ and ${M}_\pi$ are the $\sigma$ and pion masses
dressed by contributions from the diagrams of Fig.\ \ref{dbubble}, 
\begin{subequations} 
\bea \label{Msigma} 
M_{\sigma}^{2} & = & m_\sigma^2 (\sigma) +
\frac{4\lambda}{N} \left[  3\, \int_q \, {\cal
S}(q) + (N-1)\, \int_q \, {\cal P}(q) \right] \\
M_{\pi}^{2} & = & m_\pi^2(\sigma) + \frac{4\lambda}{N} \left[ 
\int_q \, {\cal S}(q)+ (N+1)\,  \int_q \, {\cal P}(q)\right] \,\, . 
\label{Mpi} 
\eea
Using Eqs.\ (\ref{Msigma}) and (\ref{Mpi}), 
Eq.\ (\ref{stationONH}) can be written
in the compact form:
\be \label{phi4} 
H = \sigma \left[M_{\sigma}^2 - \frac{8 \lambda}{N} \, \sigma^2 \right]\,\, . 
\ee
\end{subequations}
Equations (\ref{Msigma}), (\ref{Mpi}), and (\ref{phi4}) are the
stationarity conditions of the $O(N)$ model in the Hartree
approximation. The explicit calculation of the tadpole integrals
$\int_q {\cal S}(q)$ and $\int_q {\cal P}(q)$ will be discussed
in Sec.\ \ref{IId}.

\subsection{The $U(N_f)_r \times U(N_f)_\ell$ linear sigma model 
for $N_f=2, \, 3$ and $4$ flavors} \label{IIc}

The application of the CJT formalism to the 
$U(N_f)_r \times U(N_f)_\ell$ linear sigma model for $N_f = 3$ was
discussed in Ref.\ \cite{Lenaghan:2000ey}. Since we want 
to treat the cases $N_f = 2$ and $N_f=4$ on the same footing, we
derive the CJT effective potential in somewhat greater
detail than in the last subsection.

The Lagrangian of the $U(N_f)_r \times U(N_f)_\ell$ linear sigma model 
for $N_f=2,\, 3$ or 4 flavors is given by 
\cite{Levy,Hu:1974qb,Schechter:1975ju,Geddes:1980nd} 
\bea \label{LU2} 
{\cal L}(\Phi) &=& {\rm Tr} \left( \partial_{\mu} \Phi^{\dagger}
\partial^{\mu} \Phi -  m^2 \, \Phi^{\dagger} \Phi \right) - 
\lambda_{1} \left[ {\rm Tr}  \left( \Phi^{\dagger}
 \Phi  \right) \right]^{2} -
\lambda_{2} {\rm Tr}  \left( \Phi^{\dagger}  \Phi  \right)^{2} \nonumber \\
&+& c \, \left[ {\rm det} \left( \Phi \right) + {\rm det}  \left(
\Phi^{\dagger} \right) \right] + {\rm Tr} \left[H  \left(\Phi +
\Phi^{\dagger} \right)\right] \,\, . 
\eea 
$\Phi$ is a complex $N_f \times
N_f$ matrix parametrizing the scalar and pseudoscalar mesons,
\begin{subequations}
\be \Phi = T_{a} \, \phi_{a} =   T_{a} \, (\sigma_{a} +
        i \pi_{a})\,\, , \label{defphi}
\ee 
where $\sigma_{a}$ are the scalar ($J^{\rm P}=0^+$) 
fields and $\pi_{a}$ are the
pseudoscalar ($J^{\rm P}=0^-$) fields. 
The $N_f \times N_f$ matrix $H$ breaks the
symmetry explicitly and is chosen as 
\be 
H = T_{a} \, h_{a} \,\, ,
\ee
\end{subequations}
where $h_{a}$ are external fields. $T_{a}$ are the generators of
$U(N_f)$. The $T_{a}$ are normalized such that ${\rm Tr} (T_{a}
T_{b}) = \delta_{ab}/2$. They obey the $U(N_f)$ algebra with
\begin{subequations}
\bea
\left[T_{a},T_{b}\right] &=& i \, f_{abc} \, T_{c} \,\, , \\
\left\{T_{a},T_{b}\right\} &=&  d_{abc} \, T_{c} \,\, , 
\eea 
where $f_{abc}$ and $d_{abc}$ are the standard
antisymmetric and symmetric structure constants of $SU(N_f)$,
$a,b,c=1,\ldots,N_f^2-1$,  and 
\be 
f_{ab0} \equiv 0 \,\,\,\, ,\,\,\,\, d_{ab0} \equiv \sqrt{\frac{2}{N_f}}
\, \delta_{ab} \,\, . 
\ee
\end{subequations}
The terms in the first line of Eq.\ (\ref{LU2}) are invariant
under $U(N_f)_{r} \times U(N_f)_{\ell} \cong
U(N_f)_V \times U(N_f)_A$ transformations.
The determinant terms are invariant under 
$SU(N_f)_{r} \times SU(N_f)_{\ell} \cong
SU(N_f)_V \times SU(N_f)_A$, but break the $U(1)_A$ symmetry explicitly.
These terms arise from the $U(1)_A$ anomaly of the QCD vacuum.
The last term in Eq.\ (\ref{LU2}) breaks the axial and possibly the
$SU(N_f)_{V}$ vector symmetries explicitly. 

In the following we discuss the three different cases
$N_f=2, \, 3,$ and 4 in detail. The identification 
of the $\sigma_{a}$ and $\pi_a$ fields with the physical scalar 
and pseudoscalar mesons is given in Appendix \ref{appA}.

A non-vanishing vacuum expectation value for $\Phi$,
\be \label{vev}
\langle \Phi \rangle \equiv T_a \, \bar{\sigma}_{a} \,\, ,
\ee 
breaks the chiral symmetry spontaneously. (If the vacuum does
not break parity, the fields $\pi_a$ cannot
assume a non-vanishing vacuum expectation value.)
Shifting the $\Phi$ field by this expectation value,
the Lagrangian can be rewritten as
\begin{eqnarray}
{\cal L}  &=&
  \frac{1}{2} \left[ \partial_{\mu} \sigma_{a}
        \partial^{\mu} \sigma_{a} + \partial_{\mu} \pi_{a}
        \partial^{\mu} \pi_{a} - \sigma_{a} (m_{S}^{2})_{ab}
        \sigma_{b} - \pi_{a} (m_{P}^{2})_{ab}
        \pi_{b} \right]  \nonumber \\
    &-&  \, \left[ \,\frac{4}{3}\, {\cal F}_{abcd}\, \bar{\sigma}_{d}
             - \delta(N_f,3)\,{\cal G}_{abc} +
    \frac{4}{3}\, \delta(N_f,4)\, {\cal G}_{abcd}\, \bar{\sigma}_{d}\,
          \right]  \, \sigma_{a} \sigma_{b} \sigma_{c} \nonumber\\
    &-& \, \left[\frac{}{} 4\, {\cal H}_{abcd}\, \bar{\sigma}_{d}+3\,
         \delta(N_f,3)\, {\cal G}_{abc} - 4\, \delta(N_f,4)\,
         {\cal G}_{abcd}\, \bar{\sigma}_{d}\, \right]
         \,  \pi_{a} \pi_{b} \sigma_{c} \nonumber \\
        &-& 2\left[ \frac{}{} {\cal H}_{abcd}-\delta(N_f,4) \, 
          {\cal G}_{abcd}\, \right]  \, \sigma_{a} \sigma_{b}
        \pi_{c} \pi_{d} \nonumber\\
    &-& \frac{1}{3}  \, \left[ \frac{}{} {\cal F}_{abcd}+
        \delta(N_f,4)\, {\cal G}_{abcd}\, \right]  \, (
        \sigma_{a} \sigma_{b} \sigma_{c} \sigma_{d} +
        \pi_{a} \pi_{b} \pi_{c} \pi_{d} )\nonumber \\
         &-& U(\bar{\sigma}) \,\, ,
\end{eqnarray}
where $\delta(n,m) \equiv \delta_{nm}$ is the Kronecker delta
and the fields $\sigma_a$ and $\pi_a$ are the
fluctuations around the expectation values $\bar{\sigma}_a$.
The latter are determined from the condition 
$dU(\bar{\sigma})/d \bar{\sigma}_a = 0$.
The tree-level potential is
\begin{eqnarray}
\label{UtreeU2} 
U(\bar{\sigma}) &=& \frac{m^2}{2} \, \bar{\sigma}_{a}^{2} -
        \left[ \frac{}{} 3\, \delta(N_f,2)\, {\cal G}_{ab} + 
              \delta(N_f,3)\, {\cal G}_{abc}\, \bar{\sigma}_c \,\right]
              \bar{\sigma}_{a} \bar{\sigma}_{b} \nonumber\\
        &+&\frac{1}{3} \left[ \frac{}{} {\cal F}_{abcd}
           +\delta(N_f,4)\, {\cal G}_{abcd}\, \right]\,
        \bar{\sigma}_{a} \bar{\sigma}_{b} \bar{\sigma}_{c}
        \bar{\sigma}_{d} - h_{a}\, \bar{\sigma}_{a}\,\, .
\end{eqnarray}
The coefficients ${\cal F}_{abcd}$, ${\cal G}_{ab}$, ${\cal G}_{abc}$,
${\cal G}_{abcd}$, and ${\cal H}_{abcd}$ are given by
\begin{subequations}
\begin{eqnarray}
 {\cal F}_{abcd} &=&
\frac{\lambda_1}{4} \,
  \left(\delta_{ab} \delta_{cd} +
  \delta_{ad} \delta_{bc} + \delta_{ac} \delta_{bd} \right) 
        + \frac{\lambda_{2}}{8} \, \left(d_{abn} d_{ncd} +
        d_{adn} d_{nbc} + d_{acn} d_{nbd} \right) \,\, ,\\
{\cal G}_{ab} &=&
        \frac{c}{6} \left[ \frac{}{} \delta_{a0} \delta_{b0}
        - \delta_{a1} \delta_{b1}- \delta_{a2} \delta_{b2}
        - \delta_{a3} \delta_{b3} \; \right] \,\, ,  \\
 {\cal G}_{abc} &=&
     \frac{c}{6} \left[ d_{abc}-\frac{3}{2}\, (\delta_{a0}d_{0bc}
   + \delta_{b0}d_{a0c} + \delta_{c0} d_{ab0})
   + \frac{9}{2}\, d_{000}\delta_{a0} \delta_{b0} \delta_{c0}\right] \,\, , \\
 {\cal G}_{abcd} &=& - \frac{c}{16}\, \left[ \frac{}{}
        \delta_{ab} \delta_{cd} +
        \delta_{ad} \delta_{bc} + \delta_{ac} \delta_{bd} 
         - \left( d_{abn} d_{ncd} +
        d_{adn} d_{nbc} + d_{acn} d_{nbd} \right) 
+ 16\, \delta_{a0}\delta_{b0}\delta_{c0}\delta_{d0} \right. \nonumber\\
   &  &\hspace*{0.5cm} -\; 4 \left(\delta_{a0}\delta_{b0}\delta_{cd}
    +\delta_{a0}\delta_{c0}\delta_{bd} 
    +\delta_{a0}\delta_{d0}\delta_{bc}
    +\delta_{b0}\delta_{c0}\delta_{ad}
    +\delta_{b0}\delta_{d0}\delta_{ac}
    +\delta_{d0}\delta_{c0}\delta_{ab} \right) \nonumber\\
    &  & \hspace*{0.5cm} + \;
      \sqrt{8} \left. \left(\delta_{a0}d_{bcd}+\delta_{b0}d_{cda}
  +\delta_{c0}d_{dab}+\delta_{d0}d_{abc}\right) \frac{}{}\right] 
\,\, ,\\
 {\cal H}_{abcd} &=& \frac{\lambda_{1}}{4}  \, \delta_{ab} \delta_{cd}  +
     \frac{\lambda_{2}}{8} \, \left(d_{abn} d_{ncd} 
    + f_{acn} f_{nbd} + f_{bcn} f_{nad} \right) \,\, .
\end{eqnarray}
\end{subequations}
The tree-level masses, $[m_{S}^{2}]_{ab}$ and $[m_{P}^{2}]_{ab}$,
are given by
\begin{subequations}\label{massmatrices}
\begin{eqnarray}
\left[m_{S}^{2}(\bar{\sigma})\right]_{ab} &=& m^{2}  \delta_{ab} - 6   
\left[ \delta(N_f,2)\, {\cal G}_{ab}+\delta(N_f,3) {\cal G}_{abc}
\bar{\sigma}_{c} \right] +  4 \left[ {\cal F}_{abcd} 
   + \delta(N_f,4) {\cal G}_{abcd}\right] \, 
   \bar{\sigma}_{c} \bar{\sigma}_{d} \, ,\\
\left[m_{P}^{2}(\bar{\sigma})\right]_{ab} &=& m^{2}  \delta_{ab} 
   + 6   \left[ \delta(N_f,2)\, {\cal G}_{ab} 
   +\delta(N_f,3) {\cal G}_{abc} \bar{\sigma}_{c}\right]
  + 4 \left[  {\cal H}_{abcd} - \delta(N_f,4) {\cal G}_{abcd} \right]\, 
   \bar{\sigma}_{c} \bar{\sigma}_{d} \, .
\end{eqnarray}
\end{subequations}
In general, these mass matrices are not diagonal. Consequently,
the fields ($\sigma_a$, $\pi_a$) in the standard basis 
of $U(N_f)$ generators are not mass eigenstates. Since the mass
matrices are symmetric and real, diagonalization is achieved by an
orthogonal transformation, 
\begin{subequations} 
\bea \label{ortho}
\tilde{\sigma}_{i} &=& O^{(S)}_{ia} \, \sigma_{a} \,\, , \\
\tilde{\pi}_{i} &=& O^{(P)}_{ia} \, \pi_{a} \,\, , \\
\left[\tilde{m}_{S,P}^{2}\right]_{i} &=& O^{(S,P)}_{ai} \,
 \left[m_{S,P}^{2}\right]_{ab} \, O^{(S,P)}_{bi} \,\,. \label{orthoc}
\eea
\end{subequations}

The effective potential of the $U(N_f)_{r}\times U(N_f)_{\ell}$ 
linear sigma model in the CJT formalism reads
\bea 
V[\bar{\sigma},\bar{S},\bar{P}] &=&
U(\bar{\sigma}) +\half \int_k \, \left\{ \left[\ln \bar{S}^{-1}(k)\right]_{aa}
+ \left[\ln \bar{P}^{-1}(k)\right]_{aa}\right\} \nonumber \\
&+& \half\,\int_k\,   \left[ S^{-1}_{ab}(k;\bar{\sigma}) \,
\bar{S}_{ba}(k) + P^{-1}_{ab}(k;\bar{\sigma}) \, \bar{P}_{ba}(k) 
- 2 \delta_{ab} \, \delta_{ba} \right] +
V_2[\bar{\sigma},\bar{S},\bar{P}] \,\, . \label{V2U2}
\eea 
Here, $U(\bar{\sigma})$ is the tree-level potential of Eq.\ (\ref{UtreeU2}), 
and
\begin{subequations}
\bea 
S_{ab}^{-1}(k;\bar{\sigma}) &=& -k^2 \, \delta_{ab} +
        \left[m_{S}^{2}(\bar{\sigma}) \right]_{ab} \,\, , \\
P_{ab}^{-1}(k;\bar{\sigma}) &=& -k^2 \, \delta_{ab} +
        \left[m_{P}^{2}(\bar{\sigma}) \right]_{ab} \,\, ,
\eea
\end{subequations}
are the tree-level propagators for scalar and pseudoscalar
particles, with the respective mass matrices (\ref{massmatrices}).
The fluctuations $\sigma_a$ and $\pi_a$ around the expectation
values $\bar{\sigma}_a$ no longer occur in the effective
potential (\ref{V2U2}). Therefore, from now on we use the symbol
$\sigma_a$ for the expectation values for the scalar fields
in the effective potential (\ref{V2U2}).
These expectation values, and 
the full propagators for scalar, ${\cal S}(k)$, and pseudoscalar,
${\cal P}(k)$, particles are determined from the 
stationarity conditions
\begin{subequations}
\bea \label{stationphi2}\left. \frac{\delta
V[\bar{\sigma},\bar{S},\bar{P}]}{\delta
\bar{\sigma}_{a}} \right|_{\bar{\sigma}=\sigma,
\bar{S}={\cal S},\bar{P}={\cal P}} = 0 \,\, &,& \\
 \left.
\frac{\delta V[\bar{\sigma},\bar{S},\bar{P}]}{\delta
\bar{S}_{ab}} \right|_{\bar{\sigma}=\sigma,
\bar{S}={\cal S},\bar{P}={\cal P}} = 0 \,\, &,& \ \left. \frac{\delta
V[\bar{\sigma},\bar{S},\bar{P}]}{\delta \bar{P}_{ab}} 
\right|_{\bar{\sigma}=\sigma,\bar{S}={\cal S},\bar{P}={\cal P}} = 0 \,\, . 
\label{stationP} 
\eea
\end{subequations} 
With Eq.\ (\ref{selfenergy}), the latter two equations can
be written in the form
\begin{subequations}
\bea \label{schwinger}
{\cal S}_{ab}^{-1}(k) &=&
S_{ab}^{-1}(k;\bar{\sigma}) +   \Sigma_{ab}(k)\,\, , \\
{\cal P}_{ab}^{-1}(k) &=& 
P_{ab}^{-1}(k;\bar{\sigma}) +   \Pi_{ab}(k)\,\, ,
\eea
\end{subequations}
where
\begin{subequations} \label{selfenergyU2}
\bea 
\Sigma_{ab}(k) &\equiv& \left. 2 \, \frac{\delta V_2
[\bar{\sigma}, \bar{S}, \bar{P}]}{\delta \bar{S}_{ba}(k)}
\right|_{\bar{\sigma}=\sigma,\bar{S}={\cal S},\bar{P}={\cal P}} \,\, , \\
\Pi_{ab}(k) &\equiv& \left. 2 \, \frac{\delta V_2 [\bar{\sigma},
        \bar{S}, \bar{P}]}{\delta \bar{P}_{ba}(k)}
\right|_{\bar{\sigma}=\sigma,\bar{S}={\cal S},\bar{P}={\cal P}} \,\, , 
\eea
\end{subequations}
are the self-energies for the scalar and pseudoscalar particles.
As in the case of the $O(N)$ model, 
we include only the two-loop diagrams of Fig.\ \ref{dbubble} in $V_2$.
Then,
\begin{eqnarray}
 V_{2}[\bar{S}, \bar{P}] &=&
\left[ {\cal F}_{abcd}+\delta(N_f,4)\,{\cal G}_{abcd} \right] \, \left[
        \int_{k} \bar{S}_{ab}(k) \int_{p} \bar{S}_{cd}(p) +
        \int_{k} \bar{P}_{ab}(k) \int_{p} \bar{P}_{cd}(p) \right] \nonumber\\
    &+&2 \left[  {\cal H}_{abcd}-\delta(N_f,4)\,{\cal G}_{abcd}\right]
        \int_{k} \bar{S}_{ab}(k) \int_{p} \bar{P}_{cd}(p) \,\, .
\end{eqnarray}
Note that, in the Hartree approximation, $V_{2}$ is 
independent of $\bar{\sigma}_a$.  The stationarity
conditions for the condensates are
\begin{eqnarray} \label{h}
 h_{a}&=&m^2 \, \sigma_{a} - 
\left[ 6\, \delta(N_f,2)\,{\cal G}_{ab} +3\,\delta(N_f,3)\, {\cal G}_{abc}\,
 \sigma_c \right] \sigma_{b} 
        + \frac{4}{3}\, \left[{\cal F}_{abcd}+\delta(N_f,4)\,{\cal G}_{abcd}
\right]\, \sigma_{b} \sigma_{c}\sigma_{d}\nonumber\\
    &+& \left\{ - 3\, \delta(N_f,3)\,{\cal G}_{abc}
    + 4\left[ {\cal F}_{abcd} + \delta(N_f,4)\, {\cal G}_{abcd}\right]
 \sigma_{d} \right\} \int_{k}  {\cal S}_{cb}(k) \nonumber\\
    &+& \left\{ 3 \, \delta(N_f,3)\,{\cal G}_{abc}
    + 4 \left[ {\cal H}_{bcad} - \delta(N_f,4)\,{\cal G}_{abcd}\right]
\, \sigma_{d} \right\} \int_{k}  {\cal P}_{cb}(k) \,\, .
\end{eqnarray}
In the Hartree approximation, the self-energies (\ref{selfenergy}) are
independent of momentum, and the Schwinger--Dyson equations
(\ref{schwinger}) for the full propagators assume the simple form 
\bea 
{\cal S}^{-1}_{ab}(k) &=& -k^2 \, \delta_{ab} +
        \left[ M_{S}^{2} \right]_{ab} \,\, , \label{invS}\\
{\cal P}^{-1}_{ab}(k) &=& -k^2 \, \delta_{ab} +
        \left[ M_{P}^{2} \right]_{ab} \,\, . \label{invP}
\eea 
The scalar and pseudoscalar mass matrices are given by
\begin{subequations} \label{massmatrices2}
\begin{eqnarray}
\left[ M_S^2\right]_{ab}&=& \left[m_S^2(\sigma)\right]_{ab} \nonumber \\
        &+& 4 \left[ {\cal F}_{abcd}+\delta(N_f,4)\, {\cal G}_{abcd} 
             \right] \int_{q}  {\cal S}_{cd}(q)
        + 4 \left[ {\cal H}_{abcd}-\delta(N_f,4)\, {\cal G}_{abcd} 
            \right] \int_{q}  {\cal P}_{cd}(q)\,\, ,\\
\left[ M_P^2 \right]_{ab}&=&  \left[m_P^2(\sigma) \right]_{ab} \nonumber\\
        &+& 4 \left[ {\cal F}_{abcd}+\delta(N_f,4)\,{\cal G}_{abcd}\right]
              \int_{q}  {\cal P}_{cd}(q) 
        +  4 \left[ {\cal H}_{abcd}-\delta(N_f,4)\, {\cal G}_{abcd} \right]
             \int_{q}  {\cal S}_{cd}(q) \,\, .
\end{eqnarray}
\end{subequations}
In the Hartree approximation, all particles are stable
quasiparticles, i.e., the imaginary parts of the self-energies
vanish.  Therefore, the inverse propagators (\ref{invS}) and
(\ref{invP}) are real-valued.  They are also symmetric in the
standard basis of $U(N_f)$ generators and thus diagonalizable via
an orthogonal transformation.  This transformation is given by
Eq.\ (\ref{orthoc}), with the obvious replacements 
\be 
\left[ m_{S,P}^{2} \right]_{ab} \rightarrow \left[ M_{S,P}^{2}
\right]_{ab} \,\,\,\, , \,\,\,\, \left[ \tilde{m}_{S,P}^{2}
\right]_{i} \rightarrow \left[ \tilde{M}_{S,P}^{2} \right]_{i}\,\, . 
\label{replace} 
\ee
The propagator matrices are diagonalized by the same orthogonal
transformation as their inverse. The tadpole integrals
in Eqs.\ (\ref{h}) and (\ref{massmatrices2}) are therefore
computed as
\be \label{rotatedprops}
\int_q {\cal S}_{bc}(q) = O^{(S)}_{bi}  \int_q \tilde{\cal S}_i(q) \,
O^{(S)}_{ci}\,\,\,\, , \,\,\,\,\,
\int_q {\cal P}_{bc}(q) = O^{(P)}_{bi}  \int_q \tilde{\cal P}_i(q) \,
O^{(P)}_{ci} \,\,.
\ee
After this rotation, only tadpole integrals over propagators
with a single index have to be computed. This 
will be discussed in the next section.

\subsection{Explicit calculation of loop integrals and the effective
potential} \label{IId}

In principle, the calculation of the tadpole
integrals $\int_q \tilde{\cal S}_{i}(q)$, 
$\int_q {\cal S}(q)$ and $\int_q \tilde{\cal P}_{i}(q)$, and
$\int_q {\cal P}(q)$ requires renormalization. 
Renormalization of many-body approximation schemes is nontrivial,
but does not change the results qualitatively
\cite{Lenaghan:1999si,Lenaghan:2000ey}. We therefore simply omit
the vacuum contributions to the loop integrals,
\begin{subequations} \label{int1}
\bea 
\int_q \tilde{\cal S}_{i}(q) &=&  \int \frac{d^3{\bf q}}{(2 \pi)^3}\, 
    \frac{1}{\epsilon_{\bf q}[(\tilde{M}^{2}_{S})_{i}]} \,
    \left(\exp\left\{\frac{\epsilon_{\bf q}[(\tilde{M}^{2}_{S})_{i}
        ]}{T}\right\}-1 \right)^{-1} \,\, ,  \\
 \int_q \tilde{\cal P}_{i}(q) &=&  \int \frac{d^3{\bf q}}{(2 \pi)^3}\,
    \frac{1}{\epsilon_{\bf q}[(\tilde{M}^{2}_{P})_{i}]} \,
    \left(\exp\left\{\frac{\epsilon_{\bf q}[(\tilde{M}^{2}_{P})_{i}
        ]}{T}\right\}-1 \right)^{-1} \,\, , \\
 \int_q {\cal S}(q) &=& \int \frac{d^3{\bf q}}{(2 \pi)^3}\,
    \frac{1}{\epsilon_{\bf q}[M^{2}_{\sigma}]} \,
    \left(\exp\left\{\frac{\epsilon_{\bf q}[M^{2}_{\sigma}
        ]}{T}\right\}-1 \right)^{-1} \,\, , \\
 \int_q {\cal P}(q) &=& \int \frac{d^3{\bf q}}{(2 \pi)^3}\,
    \frac{1}{\epsilon_{\bf q}[M^{2}_{\pi}]} \,
    \left(\exp\left\{\frac{\epsilon_{\bf q}[M^{2}_{\pi}
        ]}{T}\right\}-1 \right)^{-1} \, .
\eea
\end{subequations}  
Here, $\epsilon_{\bf q}[M^{2}] =
\sqrt{{\bf q}^{2} + M^{2}}$ is the relativistic energy of
a quasiparticle with mass $M$ and momentum ${\bf q}$.

Now we compute the standard effective potential
$V(\bar{\sigma}) \equiv V[\bar{\sigma}, \hat{S}(\bar{\sigma}), 
\hat{P}(\bar{\sigma})]$ from Eq.\ (\ref{Veff}).
Since $V_2$ has the general structure
\be
V_2[\bar{S},\bar{P}]
= c_s \, \left[\int_k \bar{S}(k)\right]^2+c_p \,
\left[ \int_k \bar{P}(k) \right]^2+c_{sp}\,
\left[ \int_k \bar{S}(k) \right]\, \left[\int_p \bar{P}(p) \right]
\,\, ,
\ee
cf.\ Eqs.\ (\ref{V2ON}) and (\ref{V2U2}), the self-energies
(\ref{selfenergyU2}) assume the form
\begin{subequations}
\bea
\hat{\Sigma}(k;\bar{\sigma}) &=& \left. 2 \, \frac{\delta V_2
[\bar{\sigma}, \bar{S}, \bar{P}]}{\delta \bar{S}(k)}
\right|_{\bar{\sigma}=\sigma,\bar{S}=\hat{S},\bar{P}=\hat{P}} 
= 4\, c_s \int_q \hat{S}(q;\bar{\sigma}) + 2\,
c_{sp} \int_q \hat{P}(q;\bar{\sigma}) \,\, , \\
\hat{\Pi}(k;\bar{\sigma}) &=& \left. 2 \, \frac{\delta V_2
[\bar{\sigma}, \bar{S}, \bar{P}]}{\delta \bar{P}(k)}
\right|_{\bar{\sigma}=\sigma,\bar{S}=\hat{S},\bar{P}=\hat{P}} 
= 4\, c_p \int_q \hat{P}(q;\bar{\sigma}) + 2\,
c_{sp} \int_q \hat{S}(q;\bar{\sigma}) \,\, .
\eea
\end{subequations}
With these expressions one derives the identity
\be
-\frac{1}{2} \int_k \hat{\Sigma}(k;\bar{\sigma}) \, 
\hat{S}(k;\bar{\sigma}) 
-\frac{1}{2} \int_k \hat{\Pi}(k;\bar{\sigma}) \, 
\hat{P}(k;\bar{\sigma}) \equiv 
- 2\, V_2[\hat{S}(\bar{\sigma}),\hat{P}(\bar{\sigma})]\,\, .
\ee
This considerably simplifies the expressions for the standard
effective potential. For the $O(N)$ model we obtain
\begin{subequations}
\be 
V(\bar{\sigma}) =
U(\bar{\sigma}) + \half  
\int_k \, \ln \hat{S}^{-1}(k;\bar{\sigma}) + \frac{N-1}{2} 
\int_k \, \ln \hat{P}^{-1}(k;\bar{\sigma})- 
V_2[\hat{S}(\bar{\sigma}),\hat{P}(\bar{\sigma})] \,\, ,
\label{Veff2O4} 
\ee
while for the $U(N_f)_r \times U(N_f)_\ell$ model we have
\be
V(\bar{\sigma}) = 
U(\bar{\sigma}) + \half \sum_{i=0}^{N_f^2-1} 
\int_k \, \ln \hat{\tilde{S}}_{i}^{-1}(k;\bar{\sigma}) 
+ \half \sum_{i=0}^{N_f^2-1} 
\int_k \, \ln \hat{\tilde{P}}_{i}^{-1}(k;\bar{\sigma})- 
V_2[\hat{S}(\bar{\sigma}),\hat{P}(\bar{\sigma})]
\label{Veff2U2} \,\,. 
\ee
\end{subequations}
The momentum integrals in Eqs.\ (\ref{Veff2O4}) and
(\ref{Veff2U2}) require renormalization, too. As above,
we simply omit the vacuum contribution, which leads
to the following integrals \cite{Dolan:1974qd}:
\begin{subequations}
\bea
\int_k \ln \hat{\tilde{S}}_{i}^{-1}(k;\bar{\sigma}) 
&=& T\int \frac{d^3{\bf k}}{(2 \pi)^3}\,
 \ln\left( 1-\exp\left\{-\frac{\epsilon_{\bf k}
  [(\hat{\tilde{M}}^{2}_{S})_{i} ]}{T}\right\} \right) \,\, ,  \\
 \int_k \ln \hat{\tilde{P}}_{i}^{-1}(k;\bar{\sigma}) 
&=& T\int \frac{d^3{\bf k}}{(2 \pi)^3}\,
 \ln\left(1-\exp\left\{-\frac{\epsilon_{\bf k}
  [(\hat{\tilde{M}}^{2}_{P})_{i} ]}{T}\right\} \right) \,\, , \\
 \int_k \ln \hat{{\cal S}}^{-1}(k;\bar{\sigma}) 
&=&T \int \frac{d^3{\bf k}}{(2 \pi)^3}\,
 \ln\left(1-\exp\left\{-\frac{\epsilon_{\bf k}
 [\hat{M}^{2}_{\sigma} ]}{T}\right\} \right) \,\, , \\
 \int_k \ln \hat{{\cal P}}^{-1}(k;\bar{\sigma}) 
&=& T\int \frac{d^3{\bf k}}{(2 \pi)^3}\,
    \ln\left(1-\exp\left\{-\frac{\epsilon_{\bf k}
 [\hat{M}^{2}_{\pi} ]}{T}\right\} \right) \,\, .
\eea
\end{subequations}
In these equations, the tilde
denotes quantities which are diagonalized according
to Eqs.\ (\ref{orthoc}) and (\ref{rotatedprops}), respectively.
A hat denotes propagators computed according to
Eq.\ (\ref{hatG}). Masses with a hat are the mass terms appearing
in these propagators.

\section{Patterns of symmetry breaking and vacuum properties} 
\label{III}

In this section, we discuss the patterns of symmetry breaking in
the vacuum and determine the coupling constants of the models
from the vacuum properties of the mesons. 

\subsection{Patterns of symmetry breaking} \label{IIIa}

For the $O(N)$ model (with $N=4$)
we study two different patterns of symmetry breaking
(cf.\ Table \ref{table1}):
\begin{enumerate}
\item \underline{$H=0$:} For $\mu^2<0$ the $O(N)$ 
symmetry is spontaneously broken to $O(N-1)$, 
giving rise to a non-vanishing expectation
value for the $\sigma$ field and $N-1$ Goldstone bosons.
\item \underline{$H\neq 0$:} The term $- H \bar{\sigma}$ in
Eq.\ (\ref{UtreeON}) corresponds to nonzero quark masses in
the QCD Lagrangian. It breaks the $O(N)$ symmetry explicitly
to $O(N-1)$. The $N-1$ Goldstone bosons become pseudo-Goldstone bosons.
\end{enumerate}

For the $U(N_f)_r \times U(N_f)_\ell$ models with $N_f = 2$ and 3
we study the following patterns of symmetry breaking (cf.\
Table \ref{table1}):
\begin{enumerate}
\item \underline{$h_a=0,c=0$:} For $m^2<0$ the
global $SU(N_f)_V \times U(N_f)_A$ symmetry is broken to 
$SU(N_f)_V$, and $\Phi$ develops a
non-vanishing vacuum expectation value, $\langle \Phi \rangle =
 T_0 \bar{\sigma}_0$. By the
Vafa-Witten theorem \cite{Vafa:1984tf}, only the axial symmetries
can be spontaneously broken, while the vector symmetries stay
intact. In order to retain an $SU(N_f)_V$ symmetry, 
only the term proportional to $T_0$ survives in the sum over $a$
in Eq.\ (\ref{vev}) for the vacuum expectation value $\langle \Phi \rangle$.
Spontaneously breaking $U(N_f)_A$ leads to
$N_f^2$ Goldstone bosons which form a pseudoscalar,
$N_f^2$ dimensional multiplet. 
This case is referred to as the chiral limit without $U(1)_A$ anomaly.
\item \underline{$h_a=0,c\neq 0$:}
The symmetry is $SU(N_f)_V \times SU(N_f)_A$. A non-vanishing 
$\langle \Phi \rangle$
spontaneously breaks the symmetry to $SU(N_f)_V$, with the
appearance of $N_f^2-1$ Goldstone bosons which form a pseudoscalar,
$N_f^2-1$ dimensional multiplet. 
The $N_f^2$th pseudoscalar meson is no longer massless, because the $U(1)_A$
symmetry is already explicitly broken. This case is referred to as
the chiral limit with $U(1)_A$ anomaly.
\item \underline{$h_a\neq 0, c=0$:} In QCD this corresponds to
non-vanishing quark masses, but a vanishing $U(1)_A$ anomaly.
Since $\langle \Phi \rangle$ must carry the quantum numbers of the vacuum, only
the fields $\bar{\sigma}_a$ corresponding to the diagonal 
generators of $U(N_f)$ can be nonzero. The same holds for the
external fields $h_a$ which generate a non-vanishing expectation
value by explicitly breaking the $U(N_f)_A$ symmetry. For $N_f=2$ these are 
$h_0$ and $h_3$, for $N_f=3$ there is an additional field, $h_8$.
Because the masses of the up and down quarks are approximately equal, $m_u
\simeq m_d$, we restrict our study to $h_0 \neq 0$ and $h_3=0$
for all cases considered. Since the strange quark mass $m_s$
is larger than $ m_u \simeq m_d$, $h_8 \neq 0$. 
In this case the $SU(N_f)_V \times U(N_f)_A$ symmetry is 
explicitly broken to $SU(2)_V$. The latter symmetry remains intact
because $h_3 = 0$. This case is referred to as the case of explicit
chiral symmetry breaking without $U(1)_A$ anomaly.
\item \underline{$h_a\neq, c\neq0$:}
A $U(1)_A$ subgroup of the $U(N_f)_A$ symmetry is explicitly
broken by instantons. As explained above we restrict ourselves to
$h_3 = 0$. This case is referred to as the case of explicit
chiral symmetry breaking with $U(1)_A$ anomaly.
\end{enumerate}

For $N_f=4$, we only study the last case, i.e., explicit chiral
symmetry breaking with $U(1)_A$ anomaly. 
To break the $U(4)_A$ symmetry explicitly, in addition to
$h_0 \neq 0$ and $h_8 \neq 0$ we have to introduce a nonzero
value for the field $h_{15}$ corresponding to the fourth diagonal
generator of $U(4)$.
Since the charm quark mass is much larger than the light up
and down quark masses and the strange quark mass, $h_{15}$
is also much larger than either $h_0$ or $h_8$, cf.\ Table
\ref{para4}. Therefore, it does not make too much sense
to study the rather unrealistic first two cases of vanishing
quark masses. We therefore restrict our considerations
to the physical case of explicit chiral symmetry breaking
with $U(1)_A$ anomaly.

\subsection{Condensates and masses in the vacuum} \label{IIIb}

In this section we determine the parameters
of the different models from the vacuum values of the condensates
and the meson masses for the various symmetry breaking patterns 
discussed in Sec.\ \ref{IIIa}. 
For the $O(4)$, the $U(3)_{r} \times U(3)_{\ell}$, and the $U(4)_{r}
\times U(4)_{\ell}$ linear sigma model, we simply follow 
Refs.\ \cite{Lenaghan:1999si,Lenaghan:2000ey,Geddes:1980nd},
respectively. Since it has not been done previously, 
fitting the parameters of the $U(2)_r \times U(2)_\ell$
model is discussed in more detail.

For the $O(4)$ model, we have three parameters, $H$, $\lambda$, and
$\mu^2$, which are adjusted to reproduce the vacuum values for
the pion decay constant, $f_\pi$, the pion mass, $m_\pi$, and the
$\sigma$ mass, $m_\sigma$. For reasons explained below, for the
latter we choose $m_\sigma = 400$ MeV, instead of 600 MeV as in
Ref.\ \cite{Lenaghan:1999si}. The values for the parameters in the
chiral limit and with explicit symmetry breaking are listed in Table
\ref{para1}.

\begin{table}
\begin{center}
\begin{tabular}{|l|l|l|} \hline
               & Masses and decay constants & Parameter  set\\\hline
Explicit chiral& $ f_{\pi}=92.4\,{\rm MeV}$& 
                 $ H=(121.60\,{\rm MeV})^3$\\
symmetry       & $ m_{\sigma}=400\,{\rm MeV}$            & $\lambda=8.230 $\\
breaking       & $ m_{\pi}=139.5\,{\rm MeV}$ 	
                        & $\mu^2= -(225.41\,{\rm MeV})^2$ \\ \hline
Chiral limit  & $ f_{\pi}=90\,{\rm MeV}$ & $H=0$\\
                & $ m_{\sigma}=400\,{\rm MeV}$&  $\lambda=9.877 $\\
                & $ m_{\pi}=0\,{\rm MeV}$ & 
                                     $\mu^2= -(282.84\,{\rm MeV})^2$\\\hline
\end{tabular}
\end{center}
\vspace{3mm} 
\caption{ The masses and decay constants at vanishing
temperature and the corresponding parameter sets for the
$O(4)$ linear sigma model for the two symmetry breaking patterns
studied here.}
\label{para1}
\end{table}

In the $U(2)_{r} \times U(2)_{\ell}$ model, 
for all symmetry breaking patterns studied here there is
an (approximate) $SU(2)_V$ symmetry due to the (approximate) equality
of the up and down quark masses. Consequently, for all cases the 
vacuum expectation value is $\langle \Phi \rangle = T_0 \bar{\sigma}_0$.
At zero temperature, the equation for the condensate
$\sigma_0$ reads, cf.\ Eq.\ (\ref{h}),
\be
\label{h0_with_anomaly}
h_0= \sigma_0 \left[ m^2 - c + \left(\lambda_1+\frac{\lambda_2}{2} 
\right)\, \sigma_0^2 \right]. 
\ee
The PCAC relations determine the value of the condensate from the 
pseudoscalar meson decay constants, 
\be
f_a = d_{aa0} \, \sigma_0\,\, .
\ee
Since $d_{aa0} = 1$, all meson decay constants are
identical, and we obtain $\sigma_0 \equiv f_\pi$. 
The scalar mass matrix is diagonal, with the elements
\begin{subequations} \label{U2m_s}
\bea
m_\sigma^2 & \equiv & \left[m_S^2(\sigma_0)\right]_{00}
   =  m^2-c+ 3 \left(\lambda_1+\frac{\lambda_2}{2} \right)\sigma_0^2\,\, , \\
m_{a_0}^2 & \equiv & \left[m_S^2(\sigma_0)\right]_{11}
   = \left[ m_S^2(\sigma_0)\right]_{22}=\left[ m_S^2 (\sigma_0) \right]_{33}
   =   m^2+c+ \left(\lambda_1+\frac{3\lambda_2}{2}\right)\sigma_0^2 \,\, .
\eea
\end{subequations}
The pseudoscalar mass matrix is also diagonal, with the
elements 
\begin{subequations}  \label{U2m_p}
\bea
m_\eta^2 & \equiv & \left[m_P^2(\sigma_0)\right]_{00}
  = m^2 + c + \left(\lambda_1+\frac{\lambda_2}{2}\right)\sigma_0^2\,\, ,\\
m_\pi^2 & \equiv & \left[m_P^2(\sigma_0)\right]_{11}=
\left[m_P^2(\sigma_0)\right]_{22}=\left[m_P^2(\sigma_0) \right]_{33}
  = m^2 - c+ \left(\lambda_1 +\frac{\lambda_2}{2}\right)\sigma_0^2 \,\, .
\eea
\end{subequations}
Without the $U(1)_A$ anomaly, $c=0$,
the pions and the $\eta$ meson become degenerate
in mass. In the chiral limit, one then has four (instead of three)
Goldstone bosons. With the $U(1)_A$ anomaly, $c$ is 
positive, cf.\ Table \ref{para2}, and the $\eta$ meson
becomes heavier than the pion. At zero temperature, the (squared)
mass difference between the $\eta$ and the pion is determined by
the parameter $c$ characterizing the strength of the $U(1)_A$ anomaly,
$m_\eta^2 - m_\pi^2 = 2c$. Simultaneously, also the mass
difference between the $a_0$ and the $\sigma$ meson is determined by
this parameter,
$m_{a_0}^2 - m_\sigma^2 = 2c - 2\lambda_1 \sigma_0^2$. (As $\lambda_1 <0$,
cf.\ Table \ref{para2}, the second term always increases the mass
difference.)

The limit $c \rightarrow \infty$ corresponds to maximum 
explicit $U(1)_A$ symmetry breaking. In this limit, for realistic
values of the $\sigma$ meson and the pion mass (i.e., $m^2 -c = const.$),
the $\eta$ and $a_0$ mesons become infinitely heavy and are thus 
removed from the spectrum
of physical excitations. In this limit, the $U(2)_r \times U(2)_\ell$
is identical to the $O(4)$ model, where the $a_0$ and $\eta$ meson
are absent from the beginning.

With Eqs.\ (\ref{h0_with_anomaly}), (\ref{U2m_s}), and (\ref{U2m_p}),
we can determine the parameters of the model from 
the pion decay constant and the meson masses in the vacuum,
\begin{eqnarray}
\sigma_0&=&f_{\pi} \;\; , \;\;
\lambda_1=\frac{m_{\sigma}^2-m_{\pi}^2-m_{a_0}^2+m_{\eta}^2}{2f_{\pi}^2}
\;\; , \;\;
\lambda_2=\frac{m_{a_0}^2-m_{\eta}^2}{f_{\pi}^2}\,\, , \nonumber \\
m^2&=&m_{\pi}^2+\frac{m_\eta^2-m_\sigma^2}{2}\;\; , \;\; 
c=\frac{m_{\eta}^2-m_{\pi}^2}{2}  \;\; , \;\;
h_0=f_{\pi}m_{\pi}^2\,\, .
\end{eqnarray}
With the $U(1)_A$ anomaly, there are five parameters, 
$h_0, \lambda_1, \lambda_2, m^2$, and $c$, which can be unambiguously
determined from the five quantities
$f_\pi$, $m_\sigma$, $m_{a_0}$, $m_\eta$, and $m_\pi$.
Without the $U(1)_A$ anomaly, $c=0$, and $m_\eta = m_\pi$.
In this case, there are only four parameters and four quantities
from which the values of the parameters can be fixed.
The values for the parameters are listed in Table \ref{para2}.

\begin{table}
\begin{center}
\begin{tabular}{|l|l|l|} \hline
                        & Masses and decay constants & Parameter  set\\\hline
Explicit chiral &  $ f_{\pi}=92.4 \,{\rm MeV}$ & 
                   $ h_0=(121.60\,{\rm MeV})^3$\\
symmetry breaking &  $ m_{\sigma}=400\,{\rm MeV}$&  $\lambda_1=-31.03$\\
with $U(1)_A$     &  $ m_{a_0}=984.7\,{\rm MeV}$ & $\lambda_2= 78.52$ \\
anomaly   &  $ m_{\pi}=139.5\,{\rm MeV}$ & $m^2=(298.44\,{\rm MeV})^2$\\
    &$ m_{\eta}=547\,{\rm MeV}$  & $c=(374.00\,{\rm MeV})^2$ \\\hline
explicit chiral & $ f_{\pi}=92.4\,{\rm MeV}$ &  
                                    $h_0=(121.60\,{\rm MeV})^3$ \\
symmetry breaking    &  $ m_{\sigma}=400\,{\rm MeV}$
                     &$\lambda_1=-47.41$\\
without $U(1)_A$       	&  $ m_{a_0}=984.7\,{\rm MeV}$ &$\lambda_2=111.29$ \\
anomaly       		&  $ m_{\pi}=m_{\eta}=139.5\,{\rm MeV}$ & 
                                       $m^2=-(225.41\,{\rm MeV})^2$\\
       			&  & $c=0$ \\\hline
Chiral limit 	&  $ f_{\pi}=90\,{\rm MeV}$ & $h_0=0$\\
with $U(1)_A$       &$ m_{\sigma}=400\,{\rm MeV}$  &  $\lambda_1=-31.51$\\
anomaly       	& $ m_{a_0}=984.7\,{\rm MeV}$  & $\lambda_2= 82.77$ \\
       	& $ m_{\pi}=0$ & $m^2=(263.83\,{\rm MeV})^2$\\
      &  $ m_{\eta}=547\,{\rm MeV}$ & $c=(386.79\,{\rm MeV})^2$ \\\hline
Chiral limit 	&  $ f_{\pi}=90\,{\rm MeV}$ & $h_0=c=0$\\
without $U(1)_A$  &  $ m_{\sigma}=400\,{\rm MeV}$&$\lambda_1=-49.98$\\
anomaly       	&  $ m_{a_0}=984.7\,{\rm MeV}$ & $\lambda_2= 119.71$ \\
            &$ m_{\pi}=m_{\eta}=0$  &$m^2=-({282.84\rm MeV})^2$ \\\hline
\end{tabular}
\end{center}
\vspace{3mm} 
\caption{The masses and decay constants at vanishing
temperature and the corresponding parameter sets for the
$U(2)_{r} \times U(2)_{\ell}$ model for the four
symmetry breaking patterns studied here.} \label{para2}
\end{table}

For the $U(3)_r \times U(3)_\ell$ model, we follow 
Ref.\ \cite{Lenaghan:2000ey} in fitting the parameters of the model
to vacuum quantities. Our parameters differ from the ones
given in Ref.\ \cite{Lenaghan:2000ey}, since we use $m_\sigma = 400$ MeV,
and not 600 MeV. In the chiral limit, 
the number of parameters equals the number of vacuum quantities,
and one can again obtain a unique mapping between these sets of
quantities. With explicit chiral symmetry breaking, however,
there are fewer parameters than vacuum quantities. Consequently,
some meson masses are predicted rather than used as fit parameters.
The values for the parameters and the meson masses are given in Table
\ref{para3}. The vacuum quantities predicted by the fit are given in
bold-faced letters.

\begin{table}
\begin{center}
\begin{tabular}{|l|l|l|} \hline
              & Masses and decay constants & Parameter  set\\\hline
Explicit chiral &  $ f_{\pi}=92.4\,{\rm MeV}$  &
                            $h_0=(285.04\,{\rm MeV})^3$\\
symmetry beaking&  $ f_{K}=113\,{\rm MeV}$  &
                            $h_8=-(309.46\,{\rm MeV})^3$\\
with $U(1)_A$   & $ m_{\sigma}=400\,{\rm MeV}$&$\lambda_1=-5.38$ \\
anomaly      &${\bf m}_{a_0}={\bf 1024.6}\,{\rm MeV}$    & $\lambda_2=45.08$\\
 &$ {\bf m}_\kappa ={\bf 1116.2}\,{\rm MeV}$ & $m^2=(493.69\,{\rm MeV})^2$ \\
 &  $ {\bf m}_{f_0}={\bf 1188.7}\,{\rm MeV}$ & $c=4831.25\,{\rm MeV}$\\ 
                        &  $ m_{\pi}=139.5\,{\rm MeV}$ &     \\
                        &  $ m_{K}=493\,{\rm MeV}$                      &  \\
             		& $ {\bf m}_{\eta}={\bf 536.5}\,{\rm MeV}$ &\\
                        &  $ {\bf m}_{\eta'}={\bf 963.9}\,{\rm MeV}$&\\\hline
Explicit chiral & $ f_{\pi}=92.4\,{\rm MeV}$  & 
                           $h_0=(285.04\,{\rm MeV})^3$\\
symmetry breaking  &  $ f_{K}=113\,{\rm MeV}$ & 
                           $h_8=-(309.46\,{\rm MeV})^3$\\
without $U(1)_A$ &  $ m_{\sigma}=400\,{\rm MeV}$
                        & $\lambda_1=-24.13$ \\
anomaly   &  $ {\bf m}_{a_0}={\bf 844.4}\,{\rm MeV} $  & $\lambda_2=81.24$\\
  & $ {\bf m}_{\kappa}={\bf 1116.2}\,{\rm MeV}$ 
                                             & $m^2=(306.50\,{\rm MeV})^2$ \\
   		&  $ {\bf m}_{f_0}={\bf 1248.3}\,{\rm MeV}$   & $c=0$\\ 
               &  $ m_{\pi}=m_{\eta'}=139.5\,{\rm MeV}$ & \\
               &  $ m_{K}=493\,{\rm MeV}$&\\
                &  $ {\bf m}_{\eta}={\bf 630.6}\,{\rm MeV}$&\\\hline
Chiral limit  	&  $ f_{\pi}=f_{K}=90\,{\rm MeV}$ & $h_0=h_8=0$\\
with $U(1)_A$            &  $m_{\sigma}=400\,{\rm MeV}$
                            &  $\lambda_1=-17.48$\\
anomaly & $ m_{a_0}=m_{\kappa}= m_{f_0}=1225.8\,{\rm MeV}$  
                                                       &$\lambda_2=109.97$\\
  &  $  m_{\pi}=m_{K}=m_{\eta}=0$ & $m^2=(270.11\,{\rm MeV})^2$ \\
 &  $ m_{\eta'}=958\,{\rm MeV}$  & $c=6798.25\,{\rm MeV}$\\\hline   
Chiral limit &  $ f_{\pi}=f_{K}=90\,{\rm MeV}$  & $h_0=h_8=c=0$\\
without $U(1)_A$          &  $m_{\sigma}=400\,{\rm MeV}$ 
                          &  $\lambda_1=-55.25$\\
anomaly    &  $m_{a_0}=m_{\kappa}=m_{f_0}=1225.8\,{\rm MeV}$ 
                                                   & $\lambda_2=185.50$\\
           &  $m_{\pi}=m_{K}=m_{\eta}=m_{\eta'}=0 $ 
           &  $m^2=-(282.84\,{\rm MeV})^2$\\\hline
\end{tabular}
\end{center}
\vspace{3mm} \caption{The masses and decay constants at vanishing
temperature and the corresponding parameter sets for the
$U(3)_{r} \times U(3)_{\ell}$ model for the four
symmetry breaking patterns studied here. The masses in bold-faced letters
are predicted, the other masses and decay constants are used to
calculate the parameter set.} \label{para3}
\end{table}

For the $U(4)_{r} \times U(4)_{\ell}$ model we adjust the parameters
to obtain reasonable agreement between
the vacuum masses \cite{Hagiwara:2002pw} and the masses computed at tree-level,
and not the masses computed to one-loop order as in 
Ref.\ \cite{Geddes:1980nd}. 
As for the $U(3)_r \times U(3)_\ell$ case,
the number of parameters is smaller than the number of meson masses,
such that some meson masses cannot be fitted independently, but are
predicted within this approach.
We found that small values for the $\sigma$ meson, $m_\sigma \sim 400$ MeV, 
are favored, otherwise
the mass spectrum of the charmed mesons deviates too much from
the one in nature.
This is the reason why we choose a $\sigma$ meson mass
$m_\sigma = 400$ MeV also in the other cases discussed above.
The values for the 
parameters and vacuum quantities are listed in Table \ref{para4}.

\begin{table}
\begin{center}
\begin{tabular}{|l|l|l|} \hline
            & Masses and decay constants & Parameter  set\\\hline
Explicit chiral 	&$ f_{\pi}=92.4\,{\rm MeV}$ 
                  & $h_0=(917.24\,{\rm MeV})^3$\\
symmetry breaking      	&$f_{K}=113\,{\rm MeV} $ 
                  & $h_8=-(309.46\,{\rm MeV})^3 $\\
with $U(1)_A$       	&$ {\bf m}_{\sigma} = {\bf 400.6}\,{\rm MeV}$
                    &$h_{15}=-(1088.67\,{\rm MeV})^3$\\
anomaly  & $ {\bf m}_{a_0}={\bf 1052.6}\,{\rm MeV}$  & $\lambda_1=-0.12$\\
   &$ {\bf m}_{\kappa} ={\bf 1116.2}\,{\rm MeV}$  & $\lambda_2=4.85$ \\
 &  $ {\bf m}_{f_0}={\bf 1178.6}\,{\rm MeV}$& $m^2=(345.78\,{\rm MeV})^2$\\
                 &  $ {\bf m}_{D_{0}}={\bf 2370.0}\,{\rm MeV}$&$c=-1.50$\\
 &  $ {\bf m}_{D_{s0}}={\bf 2480.1}\,{\rm MeV}$&\\
 &  ${\bf m}_{\chi_{c0}}={\bf 3565.6}\,{\rm MeV}$&\\
 &  $ m_{\pi}=139.5\,{\rm MeV}$&\\
 &  $ m_{K}=493\,{\rm MeV}$  &\\
 &  $ {\bf m}_{\eta}={\bf 542.5}\,{\rm MeV}$ &\\
 &  $ {\bf m}_{\eta'}={\bf 1028}\,{\rm MeV}$&\\
 &  $ {\bf m}_{D} ={\bf 1944.1}\,{\rm MeV}$&\\
 &  $ {\bf m}_{D_s}={\bf 1899.1}\,{\rm MeV}$&\\
 &  $ {\bf m}_{\eta_c}={\bf 2129.8}\,{\rm MeV}$ &\\\hline
\end{tabular}
\end{center}
\vspace{3mm} \caption{The masses and decay constants at vanishing
temperature and the corresponding parameter sets for the
$U(4)_{r} \times U(4)_{\ell}$ model for the case of
explicit chiral symmetry breaking with $U(1)_A$ anomaly. The
masses written in bold-faced letters are predicted, the other masses
and decay constants are used to calculate the parameter set.}
\label{para4}
\end{table}

\section{Results}\label{IV}

In this section, we discuss the numerical results at nonzero
temperature for the cases listed in Table \ref{table1}.

\begin{figure}
\includegraphics{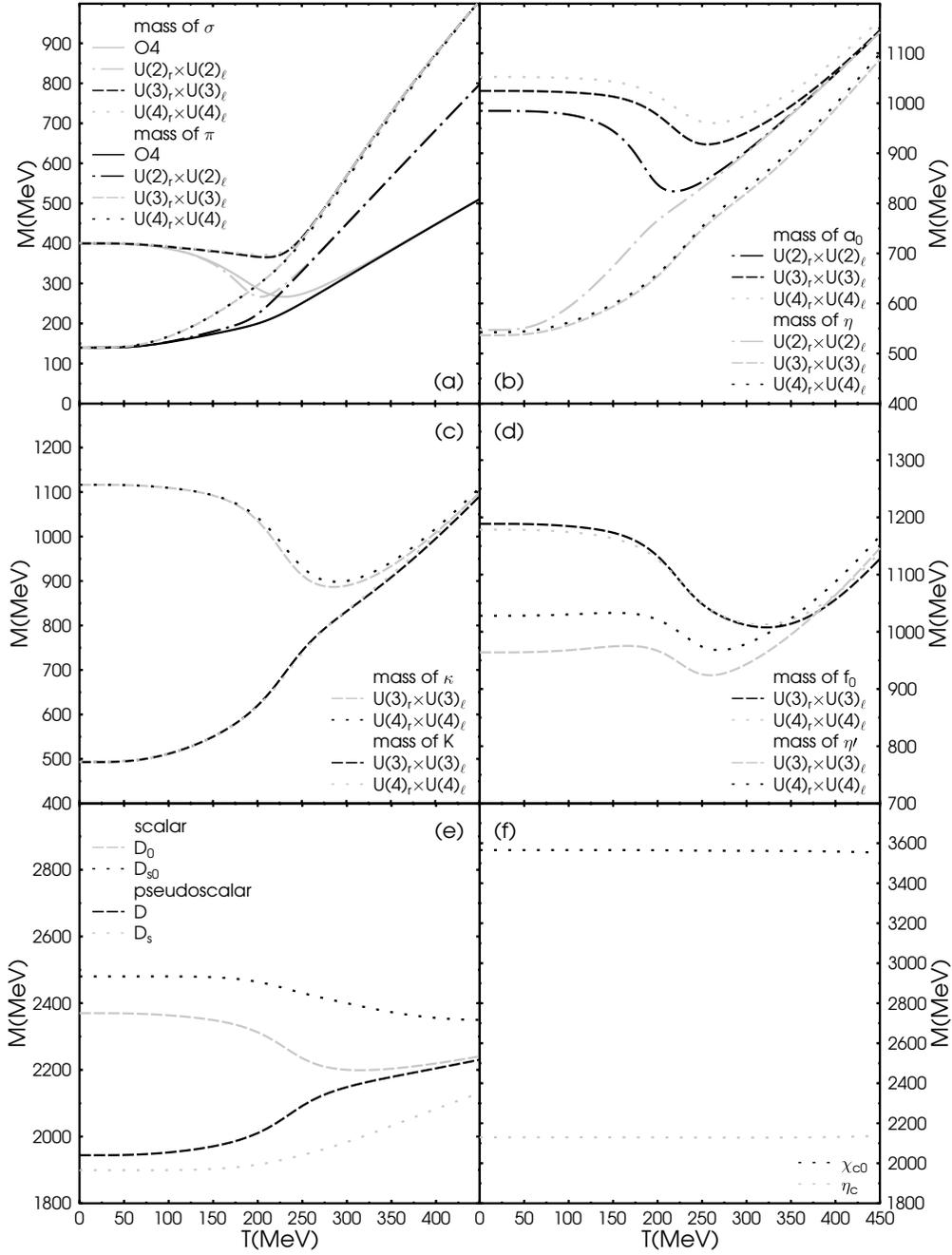}
\caption{The meson masses as a function of temperature for the
different models studied here, for the case with $U(1)_{A}$ anomaly and
explicit chiral symmetry breaking.} \label{fig1}
\end{figure}

\subsection{Explicit chiral symmetry breaking with $U(1)_{A}$ anomaly}

In Fig.\ \ref{fig1}, we show the masses of the mesons as a function
of temperature for explicit chiral symmetry breaking, including
the $U(1)_{A}$ anomaly.
This is the case where chiral symmetry breaking results in the
smallest residual symmetry group,
$SU(N_f)_V \times U(N_f)_A \rightarrow SU(2)_V$.

In Fig.\ \ref{fig1}(a) the masses of the
$\sigma$ meson and the pions are shown for all models. The
$\sigma$ meson and the pion become
degenerate in mass in the chirally restored phase. 
Comparing the $\sigma$ meson and pion masses in the $O(4)$ model
with those in the $U(2)_{r} \times U(2)_{\ell}$ model, the 
difference is almost negligible up to temperatures of 150 MeV.
In the chirally restored phase, the masses behave linearly
with temperature, but
grow faster in the $U(2)_r \times U(2)_\ell$ model than in the
$O(4)$ model. The reason is that there are twice as many fields in the
former model than in the latter, which results in twice as many
tadpole-like contributions in the equations for the in-medium masses.
These come with a positive sign and thus increase the masses.

Comparing the results of the $U(3)_{r} \times U(3)_{\ell}$ model
with those of the $U(2)_r \times U(2)_\ell$ model, one observes
differences already at a temperature of about 100 MeV. Furthermore,
the masses become even larger in the chirally restored phase.
The reason for this behavior are the strange degrees of freedom
in the $N_f=3$ case which lead to additional tadpole-like
terms in the self-energies. As above, they lead to an increase
of the in-medium masses.

Finally, one observes that virtually nothing changes
in the temperature range of interest when
including the charm degrees of freedom in the framework of the
$U(4)_r \times U(4)_\ell$ model.
The reason is that the charm quark is large compared to the temperature,
$m_c \gg T$, and the contributions from charmed particles to the
equations for the in-medium masses is suppressed.
For two reasons, this is a non-trivial result. First, the
equations for the in-medium masses are structurally different
for the $U(4)_r \times U(4)_\ell$ model as compared to the
$U(3)_r \times U(3)_\ell$ model, cf.\ Eqs.\ (\ref{massmatrices}) and
(\ref{massmatrices2}). Second, although the tadpole terms 
(\ref{int1}) are strongly suppressed for particles with masses
much larger than the temperature, Eqs.\ (\ref{h}) and 
(\ref{massmatrices2}) form a {\em nonlinear\/} system of coupled equations,
i.e., small perturbations could lead to large quantitative
changes in the solution.

In Fig.\ \ref{fig1}(b) the masses of the $a_0$ and the
$\eta$ mesons are shown as functions of temperature. 
Qualitatively, the behavior of these masses is the same
in all models.
The $a_0$ meson mass is constant up to temperatures
of 150 -- 200 MeV. It then decreases, before increasing again
above temperatures of 200 -- 250 MeV. 
The $\eta$ meson mass is constant up to $T \simeq 50$ MeV, and then
monotonously increases with temperature. 
At large temperatures, $a_0$ and $\eta$ 
become degenerate in mass, indicating restoration of chiral symmetry.
In the $U(2)_{r} \times U(2)_{\ell}$ model,
this happens somewhat earlier, at about 250 MeV, than
in the other two cases. 

In the $U(2)_r \times U(2)_\ell$ model, the $a_0$ and $\eta$ meson masses
are used to determine the parameters of the model. Thus,
at zero temperature, the masses coincide with their
correct vacuum values. In the $U(3)_r \times U(3)_\ell$
model, the predicted $\eta$ mass deviates only by 2\% from its vacuum value.
The $a_0$ mass is also rather close to the correct value;
the predicted mass is about 4\% too large.
In the $U(4)_r \times U(4)_\ell$, 
the $\eta$ mass is reproduced with excellent accuracy (the deviation
is less than 1\%),
while the $a_0$ mass is within 7\% of its vacuum value.

In Fig.\ \ref{fig1}(c) the $\kappa$ meson (now referred to
as $K_0^*(1430)$ \cite{Hagiwara:2002pw}) and kaon
masses are shown as a function of temperature. The results for
the $U(3)_{r} \times U(3)_{\ell}$ and $U(4)_{r} \times U(4)_{\ell}$
model are almost identical. The $\kappa$ meson and kaon
become degenerate in mass at temperatures of the order of 400 MeV.
In both models, the vacuum kaon mass is used as input, while the
$\kappa$ mass is predicted. The deviation to the vacuum
value in nature is about 21\%.

Figure \ref{fig1}(d) shows the masses of the
$f_0$ and the $\eta'$ meson as a function of temperature.
These mesons also become degenerate in mass at temperatures
of the order of 400 MeV.
In the $U(3)_r \times U(3)_\ell$ model, the predicted $\eta'$
mass is rather close to its value in nature; the deviation is
0.6\%.
In the $U(4)_r \times U(4)_\ell$ model, the $\eta'$ mass 
deviates from its correct vacuum value by about 7\%.
The $f_0$ mass is predicted in both models. If we
identify this state with the $f_0(1370)$, these predicted masses
deviate by 14\% from their correct values.

The masses of the $D_{s0}$, $D_0$, $D_s$, and $D$ mesons are shown
in Fig.\ \ref{fig1}(e). The
masses of the pseudoscalar mesons are known but the scalar mesons
and their masses have not yet been experimentally identified. 
It is somewhat peculiar that
the charmed, strange $D_s$ meson is lighter
than the charmed, non-strange $D$ meson. 
This is an artifact of the particular set of
coupling constants chosen here. For a different choice, this 
unphysical ordering of the masses can be reversed. Then, however,
the masses of the other mesons deviate by an unacceptably large extent
from their physical values.
Temperature has virtually no effect on the heavy charmed mesons:
their mass changes at most by 10\%, even in the chirally
restored phase. Due to the non-linear nature of
the coupled system of Eqs.\ (\ref{h}) and (\ref{massmatrices2}),
this is a non-trivial result, although not completely unexpected:
we expect significant changes of the meson masses
only when the temperature becomes of the order of the mass.
For the charmed mesons with masses of the order of 2 GeV, this
is never the case in the temperature range of interest.

Finally, in Fig.\ \ref{fig1}(f) we show the masses of
the $\chi_{c0}$ and $\eta_c$ meson. Their large masses
do not change at all for temperatures below 450 MeV.
The vacuum values for both meson masses are predicted. While the mass for the
$\chi_{c0}$ is within 5\% of its correct value, the
deviation for the $\eta_c$ is somewhat larger ($\simeq 30 \%$).

To summarize, the masses of the scalar mesons remain approximately
constant up to temperatures around 150 MeV, and then slightly decrease
before they become degenerate with the masses of the pseudoscalar mesons.
On the other hand, with the exception of the $\eta'$ mass,
the pseudoscalar masses in general increase monotonously with
temperature. 

\begin{figure}
\includegraphics{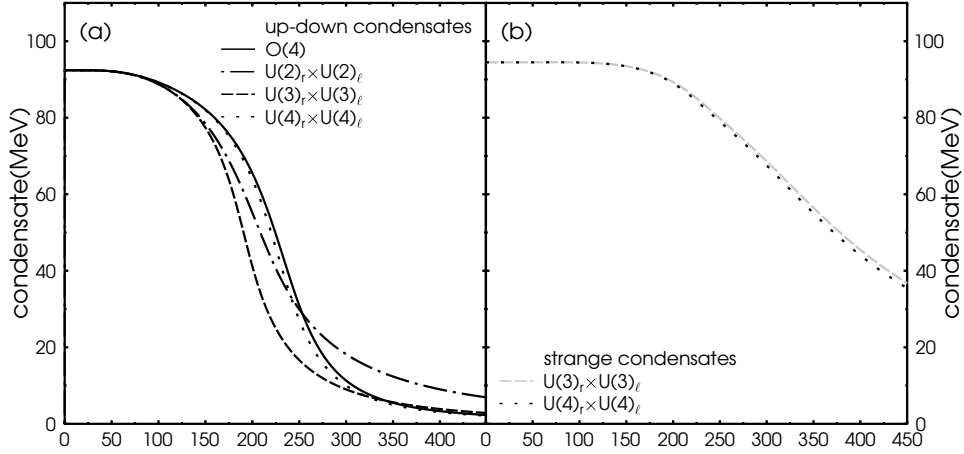}
\caption{(a) The up-down quark condensate and (b) 
the strange quark condensate as
functions of temperature for the different models in the
case with $U(1)_{A}$ anomaly and explicitly broken chiral
symmetry.} 
\label{fig2}
\end{figure}

In Fig.\ \ref{fig2}(a) the up-down quark condensate and (b) the
strange quark condensate are shown as functions of temperature.
In the $O(4)$ model and in the $U(2)_{r} \times U(2)_{\ell}$ model 
the up-down quark condensate $\varphi_{\rm up-down}$
can be directly identified with the vacuum expectation value of
the $\sigma$ field, $\varphi_{\rm up-down} \equiv \sigma_0$. 
On the other hand, in the $U(3)_{r} \times U(3)_{\ell}$ model 
\cite{Lenaghan:2000ey},
\begin{subequations} \label{condensates}
\bea 
\varphi_{\rm up-down} &=& \sqrt{\frac{2}{3}} \, \sigma_{0} +
        \frac{1}{\sqrt{3}} \, \sigma_{8}  \,\, , \\
\varphi_{\rm strange}
        &=& \frac{1}{\sqrt{3}} \, \sigma_{0} -
        \sqrt{\frac{2}{3}} \, \sigma_{8} \,\, ,
\eea
\end{subequations}
and in the $U(4)_{r} \times U(4)_{\ell}$ model
the condensates are given by
\begin{subequations}
\bea 
\varphi_{\rm up-down} &=&
        \frac{1}{\sqrt{2}} \, \sigma_{0} 
        + \frac{1}{\sqrt{3}} \, \sigma_{8} +
        \frac{1}{\sqrt{6}} \, \sigma_{15}  \,\, , \\
\varphi_{\rm strange}
        &=& \frac{1}{2} \, \sigma_{0} -
        \sqrt{\frac{2}{3}} \, \sigma_{8}  + \frac{1}{2\sqrt{3}} \,
        \sigma_{15} \, \,,        \\
\varphi_{\rm charm}
        &=& \frac{1}{2} \, \sigma_{0} -
           \frac{\sqrt{3}}{2} \, \sigma_{15} \,\, .
\eea
\end{subequations}
In these formulas $\varphi_{\rm strange}$
and $\varphi_{\rm charm}$ are the strange and
charm quark condensate, respectively.

All models predict a qualitatively similar behavior for the
temperature dependence of the up-down quark condensates. 
The strange quark condensate decreases more slowly with temperature 
than the up-down quark condensate. This is what one intuitively expects,
as it appears more difficult to ``melt'' a condensate of heavier quark
species than of light quark species. 

\begin{figure}
\includegraphics{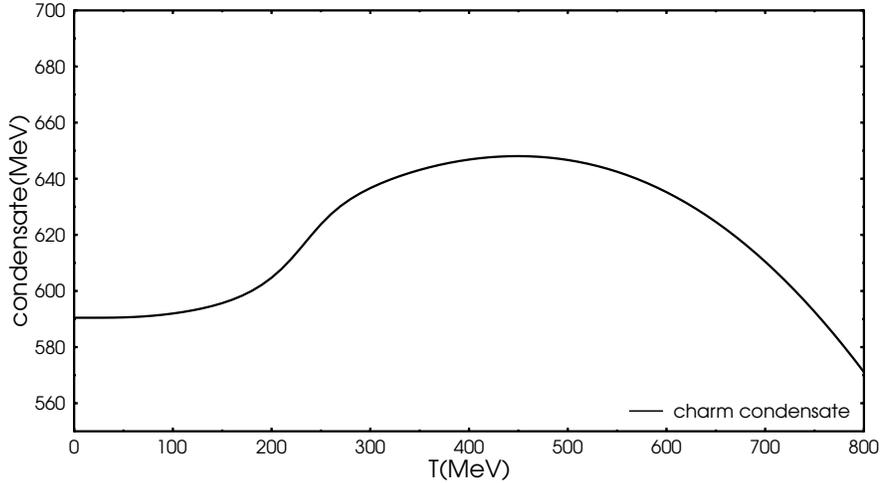}
\caption{The charm quark condensate as a function of temperature.}
\label{fig3}
\end{figure}

Figure \ref{fig3} shows the charm quark condensate.
Note that this condensate is much larger than the other
two condensates. A peculiar feature is that it first increases 
at a temperature of about 200 MeV, assumes a maximum at about
400 MeV, and then decreases. The maximum value is approximately
10\% larger than the vacuum value. We are not aware of a simple
explanation for this behavior.

\subsection{Explicit chiral symmetry breaking without $U(1)_{A}$ anomaly}

\begin{figure}
\includegraphics{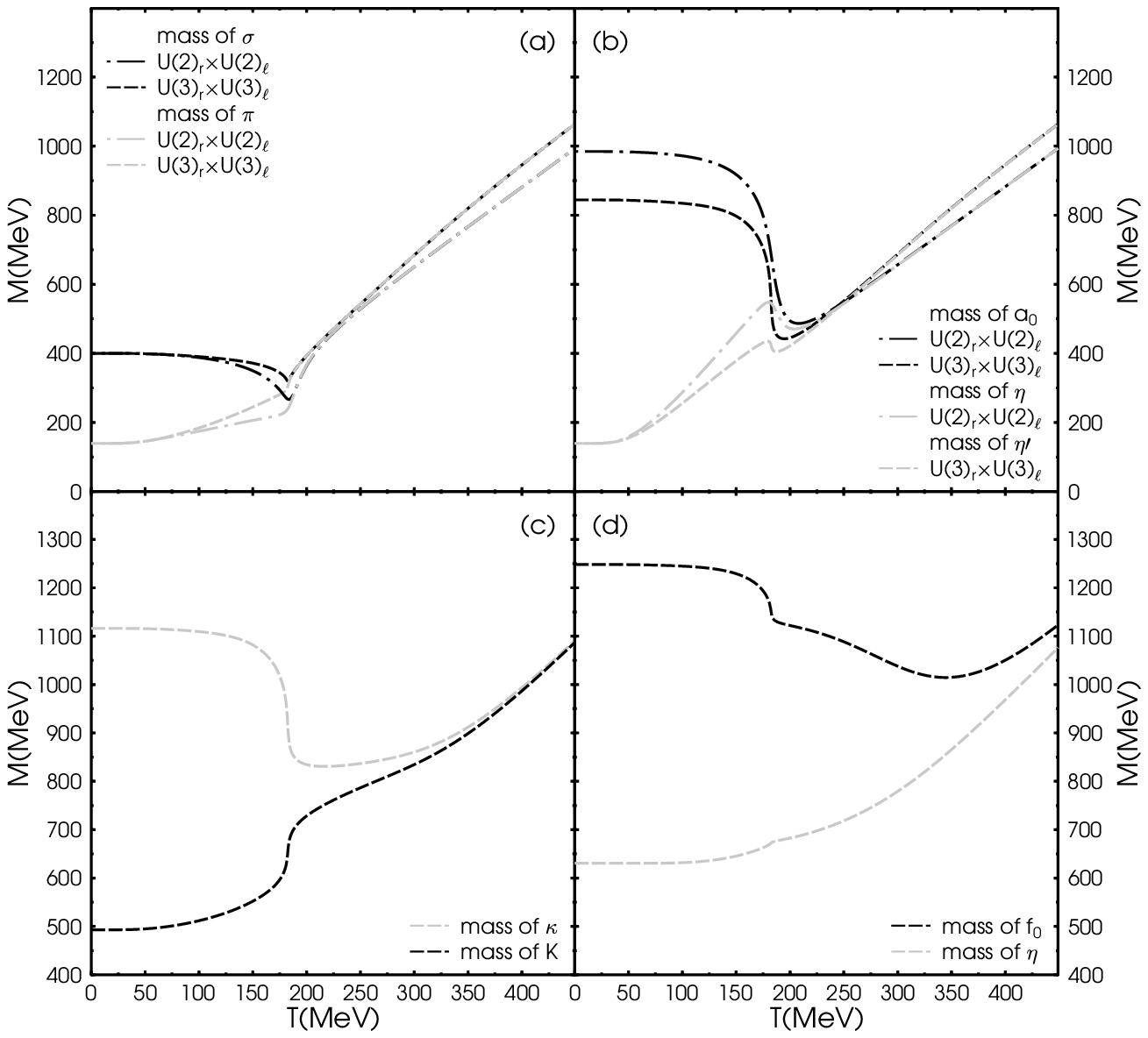}
\caption{The meson masses as a function of temperature for the
different models in the case without $U(1)_{A}$ anomaly
and explicit chiral symmetry breaking. \label{fig4}}
\end{figure}

In Fig.\ \ref{fig4}, we show the masses for the scalar and
pseudoscalar mesons for the case of explicit chiral symmetry breaking in the
absence of the $U(1)_{A}$ anomaly.
We discuss the results in comparison to the previous case. 
As in the previous case, the scalar meson masses stay constant up
to temperatures close to the transition region, then decrease and
finally start to increase again when they become degenerate with
the pseudoscalar masses. In general, the pseudoscalar masses 
increase monotonously with temperature.
The difference between the results obtained in the $U(2)_r \times U(2)_\ell$
and $U(3)_r \times U(3)_\ell$ model is rather small. An exception 
is the $a_0$ mass which is an input parameter in the former model, but
is predicted in the latter.

A marked difference to the case with $U(1)_A$ anomaly is that
the chiral symmetry restoration transition is much more rapid,
and it occurs at a slightly smaller temperature, $T \simeq 180$ MeV. 
Moreover, above the transition the scalar and pseudoscalar
masses become degenerate much more rapidly. The reason
is the absence of explicit $U(1)_A$ symmetry breaking.
At small temperatures and above the transition, the masses of the pion
and the $\eta$ meson are the same, because of the absence of the
$U(1)_{\rm A}$ anomaly. In the temperature range from about $50$
MeV to $225$ MeV, however, they are different. We believe this to be an
artifact originating from the violation of Goldstone's theorem in the
Hartree approximation, which becomes even more obvious when considering
the chiral limit.

\begin{figure}
\includegraphics{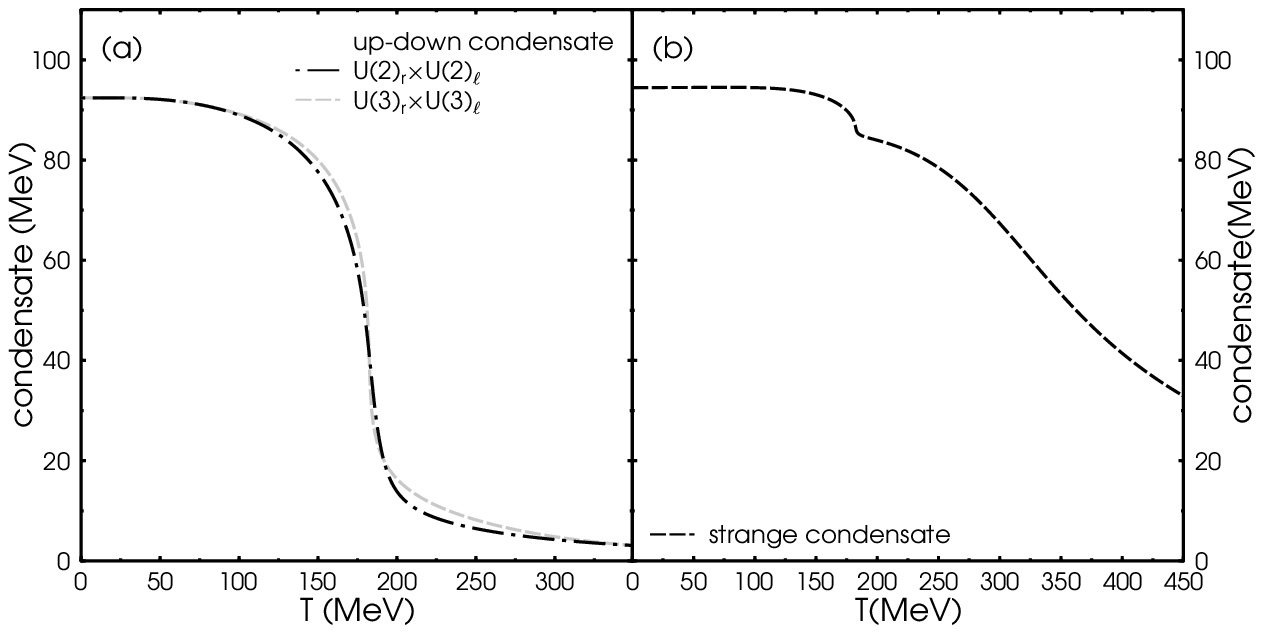}
\caption{(a) The up-down quark condensate and (b) the strange quark
condensate as a function of temperature for the different models in the
case without $U(1)_{A}$ anomaly and explicitly broken chiral
symmetry.\label{fig5}}
\end{figure}

The melting of the condensates is shown in Fig.\ \ref{fig5}.
The smaller transition temperature of about $180$ MeV is
also apparent in the temperature dependence of the
up-down quark condensate. Again, the strange quark condensate
melts less rapidly than the up-down quark condensate.

\subsection{Chiral limit with $U(1)_{A}$ anomaly}

\begin{figure}
\includegraphics{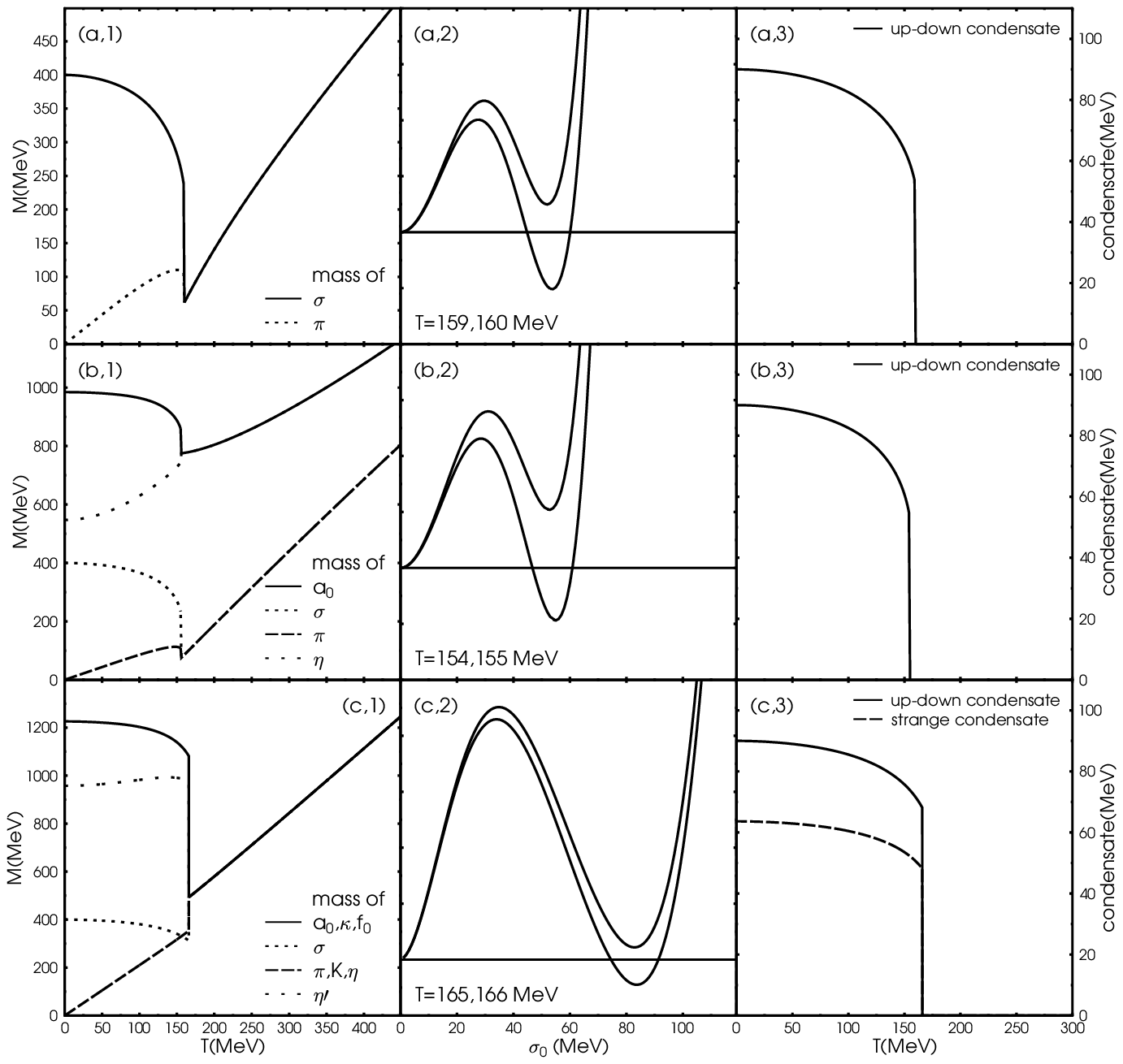}
\caption{The chiral limit with $U(1)_{A}$ anomaly. The temperature
dependence of the meson
masses are shown in panels $(~,1)$ and that of the up-down and strange 
quark condensates in panels $(~,3)$.
The results for the $O(4)$ model are shown in panels $(a,~)$,
for the $U(2)_{r} \times U(2)_{\ell}$ model in panels $(b,~)$, and
for the $U(3)_{r} \times U(3)_{\ell}$ model in panels $(c,~)$. The
effective potential (in arbitrary units) as a function
of the condensate $\sigma_0$ is shown in panels $(~,2)$.}
\label{fig6}
\end{figure}

The masses as a function of temperature for the $O(4)$ model
are shown in Fig.\
\ref{fig6}(a,1), for the $U(2)_{r} \times U(2)_{\ell}$ model
in Fig.\ \ref{fig6}(b,1), and for the $U(3)_{r} \times U(3)_{\ell}$ 
model in Fig.\ \ref{fig6}(c,1).
In the $O(4)$ and $U(2)_{r} \times U(2)_{\ell}$ models, there
are three Goldstone bosons, the pions, while in the 
$U(3)_{r} \times U(3)_{\ell}$ model there are eight Goldstone bosons,
the pions, the kaons, and the $\eta$ meson. The $\eta'$ meson
is not a Goldstone boson due to the explicit breaking
of the $U(1)_A$ symmetry by the anomaly.
The scalar octet,
comprising the three $a_0$ mesons, the four $\kappa$ mesons, and
the $f_0$ meson, is degenerate in mass, while the mass of
the singlet $\sigma$ differs from the mass of the octet. 
As the temperature increases, the
scalar masses decrease while the pseudoscalar masses increase. The
mass of the Goldstone bosons increases, because the Hartree
approximation does not respect Goldstone's theorem at nonzero
temperature \cite{Lenaghan:1999si,Lenaghan:2000ey}. 

Due to the restoration of chiral symmetry above the transition
temperature $T_c$, the
masses of the chiral partners become degenerate for temperatures
$T > T_c$. For the $O(4)$ model,
the chiral partners are the $\sigma$ and the pion, for the
$U(2)_r \times U(2)_\ell$ model they are the $\sigma$ and the pion,
as well as the $a_0$ and the $\eta$.
Due to the explicit breaking of the $U(1)_A$ symmetry, the
$\sigma/$pion and $a_0/\eta$ do not become degenerate.
The reason for this behavior is
the term $\sim {\cal G}_{ab}$ in
Eqs.\ (\ref{massmatrices}). As discussed in Sec.\ \ref{IIIb},
at zero temperature this term leads to a difference in the masses
(squared) of $\eta$ meson and pion, and of
the $a_0$ and $\sigma$ meson, respectively, which is proportional to $2c$.
In the case with $U(1)_A$ anomaly, $2c \neq 0$ even for temperatures
above $T_c$.
Consequently, these mass differences persist also in the chirally
restored phase.

For $N_f = 3$, the situation is different. The term
$\sim {\cal G}_{ab}$ for $N_f=2$ is replaced by a term 
$\sim {\cal G}_{abc} \sigma_c$. In the chirally symmetric phase,
$\sigma_c =0$, and consequently this term vanishes.
Therefore, all meson masses become degenerate, even in the case
with $U(1)_A$ anomaly.

All models exhibit
a first order phase transition between
the low-temperature phase where chiral symmetry is broken and the
high-temperature phase where chiral symmetry is restored. 
For the $O(4)$ model and the $U(2)_r \times U(2)_\ell$ model with
explicit breaking of the $U(1)_A$ symmetry, 
the transition should be second order, cf.\ Sec.\ \ref{I}.
It is a known shortcoming of the Hartree approximation to predict
a first order transition also in these cases.
For the $U(2)_r \times U(2)_\ell$ model without explicit
breaking of the $U(1)_A$ symmetry, and for all
$U(N_f)_r \times U(N_f)_\ell$ models with $N_f > 2$ the
transition is of first order, which is correctly reproduced
by the Hartree approximation.

We have determined the numerical value of $T_c$ 
by computing the effective potentials.  The latter are shown for the
three different models in the second column of Fig.\ \ref{fig6}
as a function of the condensate $\sigma_0$.
(In the chiral limit, $\sigma_0$ is the only non-trivial
condensate.)
For the extraction of the transition temperature,
the absolute normalization of the effective potential
is irrelevant. All that matters is
to identify the temperature where the minimum at the origin
and the one at a nonzero value of $\sigma_0$ become degenerate.
For this purpose we have plotted the effective potential
for two temperatures, one slightly below and one slightly above $T_c$. 
From this we deduce that, for the $O(4)$ model, Fig.\ \ref{fig6}(a,2), 
$T_c$ is between 159 and 160 MeV. For the $U(2)_{r} \times
U(2)_{\ell}$ model, Fig.\ \ref{fig6}(b,2), we obtain a critical
temperature between 154 and 155 MeV. This temperature is rather
close to the one in the $O(4)$ model.
Finally, for the $U(3)_{r} \times U(3)_{\ell}$ model,
the critical temperature is between 165 and 166 MeV, which is
slightly larger than in the previous cases.

These values are surprisingly close to those obtained from
lattice QCD calculations \cite{Karsch:2001cy}. In the chiral limit,
these calculations predict $T_c \simeq 175$ MeV for $N_f = 2$
and $T_c \simeq 155$ MeV for $N_f = 3$. The 
critical temperature obtained from the chiral models in the
Hartree approximation deviates from these values only by
about 20 MeV (or 12\%) for $N_f=2$ and only 10 MeV (or 6\%) for
$N_f = 3$. However, for the chiral models the transition temperature
is larger in the three-flavor case than in the two-flavor case,
while one finds the opposite behavior in the lattice QCD calculations.

For the sake of completeness we also show the condensates 
in the third column of Fig.\ \ref{fig6}. Note that, in contrast to
the cases where chiral symmetry is explicitly broken,
the strange condensate is smaller than the up-down condensate.
The reason is that, in the chiral limit, $\sigma_8 = 0$, such
that $\varphi_{\rm up-down} = \sqrt{2} \, \varphi_{\rm strange}$,
cf.\ Eq.\ (\ref{condensates}).

\subsection{Chiral limit without $U(1)_{A}$ anomaly}

\begin{figure}
\includegraphics{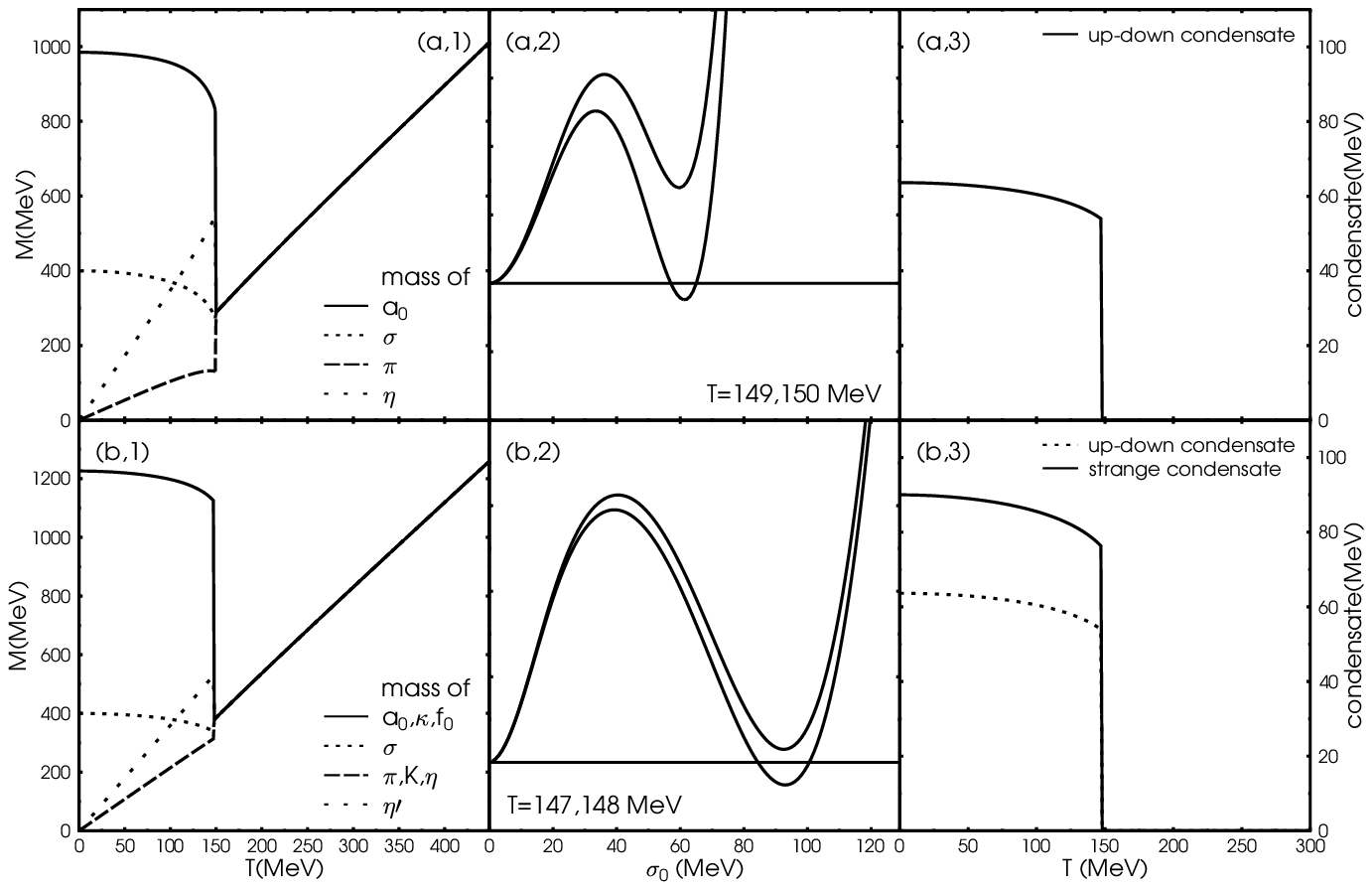}
\caption{The chiral limit without $U(1)_{A}$ anomaly.
The temperature dependence of the meson
masses are shown in panels $(~,1)$ and that of the up-down and strange 
quark condensates in panels $(~,3)$. The results 
for the $U(2)_{r} \times U(2)_{\ell}$ model in panels $(a,~)$, and
for the $U(3)_{r} \times U(3)_{\ell}$ model in panels $(b,~)$. The
effective potential (in arbitrary units) as a function
of the condensate $\sigma_0$ is shown in panels $(~,2)$.}
\label{fig7}
\end{figure}

The masses as a function of temperature are shown in Fig.\
\ref{fig7}(a,1) for the $U(2)_{r} \times U(2)_{\ell}$ model, and
in Fig.\ \ref{fig7}(b,1) for the $U(3)_{r} \times U(3)_{\ell}$ model.
In the first case, there are four Goldstone bosons,
the pions and the $\eta$ meson, while in the latter case there are
nine Goldstone bosons, the pions, the kaons, and the
$\eta$ and $\eta'$ mesons.
As the temperature increases, the scalar masses
decrease while the pseudoscalar masses increase, until they become
degenerate in a first order phase transition. The mass of the
Goldstone bosons increases with temperature, because the
Hartree approximation does not respect Goldstone's theorem at
nonzero temperature \cite{Lenaghan:1999si,Lenaghan:2000ey}. 
Moreover, the masses of the Goldstone bosons are not equal:
for the $U(2)_{r} \times U(2)_{\ell}$ model, the $\eta$ meson becomes
heavier than the pion, while for the $U(3)_{r} \times U(3)_{\ell}$ model
the $\eta'$ meson becomes heavier than the other Goldstone
bosons (pions, kaons, $\eta$ meson). Note
that the role of the $\eta$ meson in the two-flavor case is assumed
by the $\eta'$ meson in the three-flavor case. The reason is
that the physical meson corresponding to the singlet representation in
the $U(2)_r \times U(2)_\ell$ model is the
$\eta$ meson, while it is the $\eta'$ in the $U(3)_r \times U(3)_\ell$ model.

The numerical values for the critical temperature $T_c$ have been
determined by computing the effective potentials.
The latter are shown in the
second column of Fig.\ \ref{fig7} as a
a function of the condensate $\sigma_0$, for temperatures
slightly below and above $T_c$. In the $U(2)_{r} \times
U(2)_{\ell}$ model, Fig.\ \ref{fig7}(a,2) we obtain a critical
temperature of $\simeq 150$ MeV. For
the $U(3)_{r} \times U(3)_{\ell}$ model the temperature is slightly
smaller, between 147 and 148 MeV. 
The ordering of critical temperatures for the
two- and three-flavor cases is now reversed as compared to the case
with $U(1)_A$ anomaly, and consequently
in agreement with the ordering found in lattice QCD calculations.
One might be tempted to view this as an indication for
a rapidly decreasing $U(1)_A$ anomaly near the critical temperature,
in agreement with the results of Ref.\
\cite{Alles:1997nm,Schaffner-Bielich:1999uj}.
Finally, the condensates are shown in the third column
of Fig.\ \ref{fig7}. The results are qualitatively similar to the
case with $U(1)_A$ anomaly.

\section{Conclusions} \label{V}

In this work, we have used
several different chiral models, the $O(4)$, the $U(2)_{r}
\times U(2)_{\ell}$, the $U(3)_{r} \times U(3)_{\ell}$, and the
$U(4)_{r} \times U(4)_{\ell}$ linear sigma model, to
compute the temperature dependence of meson masses and quark condensates
across the chiral phase transition.
The meson masses and condensates were self-consistently
calculated in the Hartree approximation, which we derived
via the CJT formalism. Moreover, we have studied several distinct
patterns of symmetry breaking within the different models. For 
a list of cases studied here see Table \ref{table1}.

We first considered the physically relevant case of
explicit symmetry breaking in the presence of the $U(1)_A$ anomaly
and compared the results of the different chiral models
in order to clarify how they change with the number of quark flavors $N_f$.
Comparing the $O(4)$ model with
the $U(2)_{r} \times U(2)_{\ell}$ model, one first notices that
the degrees of freedom have doubled: in addition to the
$\sigma$ meson and the pions which are already present in the $O(4)$ model, 
one now has in addition the $\eta$ meson and the $a_0$ mesons.
This has the consequence that the meson masses
grow more rapidly with temperature in the phase where
chiral symmetry is restored. The reason for this are the 
tadpole contributions from the additional degrees of freedom 
to the meson self-energies, which lead to an increase in the meson masses.
This result also applies when adding the strange degree of freedom
in the framework of the $U(3)_{r} \times U(3)_{\ell}$ model.
In fact, this picture holds in general, 
as long as the masses of the additional
degrees of freedom are of the same order of magnitude as the chiral
phase transition temperature. On the other hand,
adding the heavy charm quark degree of freedom in the framework
of the $U(4)_{r} \times U(4)_{\ell}$ model
does not significantly influence the results
for the masses of the non-charmed mesons and the non-charmed condensates.
The reason is that the additional tadpole contributions from
the heavy charmed mesons are exponentially suppressed
with the meson mass, $\sim \exp(-M/T)$.
Vice versa, also the masses of the charmed mesons do not change 
appreciably from their vacuum values over the range of temperatures
of interest for chiral symmetry restoration, simply because
the tadpole contributions from the non-charmed meson are small
compared to the large vacuum mass of the charmed mesons.
This result is intuitively clear from the physical point of view, 
but is still non-trivial: first, the
equations for the in-medium masses are structurally different
for the $U(4)_r \times U(4)_\ell$ model as compared to the
$U(3)_r \times U(3)_\ell$ model. Second, 
the set of coupled equations for the masses and condensates is
a nonlinear system of equations, which means that
small perturbations could lead to large quantitative
changes in the solution.

We then studied the case of explicit chiral symmetry breaking
without $U(1)_A$ anomaly. The main difference to the previous
case was that the region of the chiral transition is narrower and
located at a somewhat smaller temperature.

Finally, we considered the meson masses and quark 
condensates in the chiral limit.
The Hartree approximation correctly predicts the chiral
transition to be of first order in the $U(2)_r \times U(2)_\ell$
model without $U(1)_A$ anomaly and in the $U(3)_r \times U(3)_\ell$
model. For the $O(4)$ model and the $U(2)_r \times U(2)_\ell$
model with $U(1)_A$ anomaly the Hartree approximation incorrectly
produces a first order instead of a second order phase transition.
The transition temperatures are surprisingly close to the ones
obtained in lattice QCD calculations. However, in the case
with $U(1)_A$ anomaly the transition temperature increases with
the number of flavors, while in lattice QCD it decreases. This
picture changes in the case with $U(1)_A$ anomaly, where the
transition temperature shows the same behavior with the number
of quark flavors as in lattice QCD. This may indicate
that the $U(1)_A$ symmetry is, at least partially, restored
at and above the chiral phase transition temperature.

\begin{acknowledgments}
J.R.\ acknowledges support from the Studienstiftung des
deutschen Volkes (German National Merit Foundation).
\end{acknowledgments}

\appendix

\section{Scalar and pseudoscalar meson fields} \label{appA}

For $N_f=2$, the identification of the physical
scalar and pseudoscalar meson fields with
the matrix fields defined in Eq.\ (\ref{defphi}) is
\begin{subequations}
\begin{eqnarray}
 T_a \sigma_a & = &
\frac{1}{\sqrt{2}} \left( \begin{array}{ccc}
 \frac{1}{\sqrt{2}}\sigma_0+\frac{1}{\sqrt{2}}a_0^0 & a_0^+\\
 a_0^- & \frac{1}{\sqrt{2}}\sigma_0-\frac{1}{\sqrt{2}}a_0^0 \\
\end{array} \right)\,\, , \\
T_a \pi_a&=& \frac{1}{\sqrt{2}} \left( \begin{array}{ccc}
 \frac{1}{\sqrt{2}}\pi_0+\frac{1}{\sqrt{2}}\pi^0 & \pi^+\\
 \pi^- & \frac{1}{\sqrt{2}}\pi_0-\frac{1}{\sqrt{2}}\pi^0 \\
\end{array} \right)\,\, .
\end{eqnarray}
\end{subequations}
Here, $\pi^{\pm}\equiv (\pi_1 \mp i \pi_2)/\sqrt{2}$ and $\pi^0
\equiv \pi_3$ are the charged and neutral pions, respectively.
Note the change of sign in the definition of the charged pion
fields $\pi^\pm$ in terms of $\pi_{1,2}$ in comparison to Ref.\
\cite{Lenaghan:2000ey}. The definition given here is the
correct one, as one can readily confirm by writing the meson fields in
terms of their quark content,
\be
\pi_a \sim \bar{q}\, T_a \, \gamma_5 \, q \,\, .
\ee
This applies also to the other charged meson fields defined in the
following.
The field $\eta \equiv \pi_0$ can be identified with the $\eta$ meson. The
parity partner of the pion is the $a_0$(980) meson, i.e.,
$a_0^{\pm}\equiv (\sigma_1 \mp i \sigma_2)/\sqrt{2}$ and $a_0^0
\equiv \sigma_3$. The field $\sigma_0$ corresponds to 
the $\sigma$ meson [now also referred to as $f_0(400-1200)$]. 

For $N_f=3$ we obtain the following matrix:
\begin{subequations}
\begin{eqnarray}
T_a\sigma_a & = & \frac{1}{\sqrt{2}} \left(
\begin{array}{ccc}
 \frac{1}{\sqrt{2}} a_0^0+\frac{1}{\sqrt{6}}\sigma_8
 +\frac{1}{\sqrt{3}}\sigma_0 & {a_0^+} & \kappa^+\\
 a_0^- & -\frac{1}{\sqrt{2}}a_0^0+\frac{1}{\sqrt{6}}\sigma_8
 +\frac{1}{\sqrt{3}} \sigma_0 & \kappa^0\\
 \kappa^- & \bar{\kappa^0} & -\frac{2}{\sqrt{3}}\sigma_8
 +\frac{1}{\sqrt{3}}\sigma_0\\
\end{array} \right)\,\, , \\
T_a \pi_a & = & \frac{1}{\sqrt{2}} \left(
\begin{array}{ccc}
 \frac{1}{\sqrt{2}}\pi^0+\frac{1}{\sqrt{6}}\pi_8
 +\frac{1}{\sqrt{3}}\pi_0 & \pi^+ & K^+\\
 \pi^- & -\frac{1}{\sqrt{2}}\pi^0+\frac{1}{\sqrt{6}}\pi_8
 +\frac{1}{\sqrt{3}}\pi_0 & K^0\\
 K^- & \bar{K^0} & -\frac{2}{\sqrt{3}}\pi_8+\frac{1}{\sqrt{3}}\pi_0\\
\end{array} \right)\,\, .
\end{eqnarray}
\end{subequations}
The fields $K^{\pm} \equiv (\pi_{4} \mp i \,
\pi_{5})/\sqrt{2}$, $K^{0} \equiv (\pi_{6} - i \,
\pi_{7})/\sqrt{2}$, and $\bar{K}^{0} \equiv (\pi_{6} + i \,
\pi_{7})/\sqrt{2}$ are the kaons.  In general, because the strange
quark is much heavier than the up or down quarks, the $\pi_{0}$
and the $\pi_{8}$ are admixtures of the $\eta$ and the $\eta'$
meson.
We identify the parity partner of the kaon
with the $\kappa$ meson [now referred to as $K_{0}^{*}(1430)$ in
\cite{Hagiwara:2002pw}]. Finally, in general the $\sigma_{0}$ and
the $\sigma_{8}$ are admixtures of the $\sigma$ and $f_{0}(980)$ mesons. 

For $N_f=4$ the following identification of physical fields
with matrix elements holds:
\begin{eqnarray}
T_a\sigma_a&=&\frac{1}{\sqrt 2} \left( \begin{array}{cccc}
 {A_S} & {a_0^+} & {\kappa^+} & {\bar D^0_0}\\
 {a_0^-} & {B_S} & {\kappa^0} & {D^-_0}\\
 {\kappa^-} & {\bar\kappa^0} & {C_S}&{D^-_{s,0}}\\
 {D^0_0} & {D^+_0} & {D^+_{s,0}} & {D_S}\\
\end{array} \right) \,\, ,
\end{eqnarray}
where
\begin{eqnarray}
A_S&=&\frac{1}{2}\sigma_0+\frac{1}{\sqrt{2}}a_0^0+\frac{1}{\sqrt{6}}\sigma_8
   +\frac{1}{\sqrt{12}}\sigma_{15}\,\, ,\nonumber\\
B_S&=&\frac{1}{2}\sigma_0-\frac{1}{\sqrt{2}}a_0^0+\frac{1}{\sqrt{6}}\sigma_8
+\frac{1}{\sqrt{12}}\sigma_{15}\,\, ,\nonumber\\
C_S&=&\frac{1}{2}\sigma_0-\frac{2}{\sqrt{6}}\sigma_8
+\frac{1}{\sqrt{12}}\sigma_{15}\,\, ,\nonumber\\
D_S&=&\frac{1}{2}\sigma_0-\frac{3}{\sqrt{12}}\sigma_{15}\,\, ,\nonumber
\end{eqnarray}
and
\begin{eqnarray}
T_a\pi_a &=&\frac{1}{\sqrt 2} \left( \begin{array}{cccc}
 {A_P} & {\pi^+} & {K^+} & {\bar D^0}\\
 {\pi^-} & {B_P} & {K^0} & {D^-}\\
 {K^-} & {\bar K^0} & {C_P}&{D^-_s}\\
 {D^0} & {D^+} & {D^+_s} & {D_P}\\
\end{array} \right) \,\, ,
\end{eqnarray}
with
\begin{eqnarray}
A_P&=&\frac{1}{2}\pi_0+\frac{1}{\sqrt{2}}\pi^0+\frac{1}{\sqrt{6}}\pi_8
+\frac{1}{\sqrt{12}}\pi_{15}\,\, ,\nonumber\\
B_P&=&\frac{1}{2}\pi_0-\frac{1}{\sqrt{2}}\pi^0+\frac{1}{\sqrt{6}}\pi_8
+\frac{1}{\sqrt{12}}\pi_{15}\,\, ,\nonumber\\
C_P&=&\frac{1}{2}\pi_0-\frac{2}{\sqrt{6}}\pi_8
+\frac{1}{\sqrt{12}}\pi_{15}\,\, ,\nonumber\\
D_P&=&\frac{1}{2}\pi_0-\frac{3}{\sqrt{12}}\pi_{15}\,\, .\nonumber
\end{eqnarray}
Here, $D^0 = (\pi_9+i\pi_{10})/\sqrt{2}$, $\bar{D}^0 = 
(\pi_9-i\pi_{10})/\sqrt{2}$, $D^\pm = 
(\pi_{11}\pm i\pi_{12})/\sqrt{2}$, and $D^\pm_s =
(\pi_{13}\pm i\pi_{14})/\sqrt{2}$ are the charged and neutral
pseudoscalar mesons with charm quantum numbers. 
The $\pi_0$, $\pi_8$, and $\pi_{15}$ fields are
admixtures of the $\eta$, $\eta'$, and $\eta_c$ mesons. The
charged and neutral scalar mesons with charm quantum numbers are $D_0^0 =
(\sigma_9+i\sigma_{10})/\sqrt{2}$, $\bar{D}_0^0 = 
(\sigma_9-i\sigma_{10})/\sqrt{2}$, $D_0^\pm = 
(\sigma_{11}\pm i\sigma_{12})/\sqrt{2}$, and $D_{s,0}^\pm = 
(\sigma_{13}\pm i\sigma_{14})/\sqrt{2}$. These scalar mesons
have not been identified experimentally, yet.
The $\sigma_0$, $\sigma_8$, and $\sigma_{15}$ fields
are admixtures of the $\sigma$, $f_0$, and $\chi_{c0}$ mesons.

\bibliography{u4}

\end{document}